\documentclass[namedreferences]{solarphysics}

\usepackage[hyperref,optionalrh]{spr-sola-addons} 
\usepackage{graphicx}		
\usepackage{color}		   
\usepackage[usenames,dvipsnames,svgnames,table]{xcolor}
\usepackage{subfig}
\usepackage{epstopdf}
\usepackage{cases}

\renewcommand{\vec}[1]{{\mathbfit #1}}

\chardef\us=`\_

\bibpunct[,]{(}{)}{,}{a}{,}{,}

\begin{document}
\begin{article}
\begin{opening}

\title{Can Multi-Threaded Flux Tubes in Coronal Arcades Support a Magnetohydrodynamic Avalanche?}

\author[addressref={aff1,aff2},corref,email={j.threlfall@abertay.ac.uk},]{\inits{J.}\fnm{J.}~\lnm{Threlfall}\orcid{0000-0001-6690-0923}}
\author[addressref=aff2,email={jr93@st-andrews.ac.uk}]{\inits{J.}\fnm{J.}~\lnm{Reid}\orcid{0000-0002-6391-346X}}
\author[addressref=aff2,email={awh@st-andrews.ac.uk}]{\inits{A.W.}\fnm{A.W.}~\lnm{Hood}\orcid{0000-0003-2620-2068}}

\address[id=aff1]{Division of Computing and Mathematics, Abertay University, Kydd Building, Dundee, DD1 1HG, UK}
\address[id=aff2]{School of Mathematics and Statistics, Mathematical Institute, University of St Andrews, St Andrews, KY16 9SS, UK}

\runningauthor{J. Threlfall et~al.}
\runningtitle{MHD avalanches in multi-threaded, toroidal flux tubes}

\begin{abstract}
Magnetohydrodynamic (MHD) instabilities allow energy to be released from stressed magnetic fields, commonly modelled in cylindrical flux tubes linking parallel planes, but, more recently, also in curved arcades containing flux tubes with both footpoints in the same photospheric plane.
Uncurved cylindrical flux tubes containing multiple individual threads have been shown to be capable of sustaining an MHD avalanche, whereby a single unstable thread can destabilise many.
We examine the properties of multi-threaded coronal loops, wherein each thread is created by photospheric driving in a realistic, curved coronal arcade structure (with both footpoints of each thread in the same plane).
We use three-dimensional MHD simulations to study the evolution of single- and multi-threaded coronal loops, which become unstable and reconnect, while varying the driving velocity of individual threads.
Experiments containing a single thread destabilise in a manner indicative of an ideal MHD instability and consistent with previous examples in the literature.
The introduction of additional threads modifies this picture, with aspects of the model geometry and relative driving speeds of individual threads affecting the ability of any thread to destabilise others.
In both single- and multi-threaded cases, continuous driving of the remnants of disrupted threads produces secondary, aperiodic bursts of energetic release.
\end{abstract}

\end{opening}

\section{Introduction} \label{sec:intro} 

Despite incredible advances, in both observations and modelling, our understanding of the exact means through which magnetic energy is released in the solar corona remains limited.
Solutions to the coronal-heating problem likely rely upon a combination of processes \citep{review:ParnellDeMoortel2012}.
One of the strongest candidates for dissipating magnetic energy in the solar corona is reconnection \citep{book:PriestForbes}, which releases energy stored in coronal magnetic structures.
This idea lies at the heart of the \lq\lq nanoflare\rq\rq\ scenario \citep{paper:Parker1988}, in which the corona is heated by many small, yet frequent, impulsive releases of magnetic energy in localized reconnection events.
Such nanoflares are extremely challenging to probe, or, indeed, even to observe on such small scales, leaving the community reliant on models underpinned by limited observational evidence.

Magnetic fields that permeate the active solar corona often emerge as, or subsequently form, distinct, cylindrical magnetic structures: flux tubes.
Twisted flux tubes are commonly associated with bursts of energy released during flares, which attest to their destabilisation and to energy transfer to local heating, waves, and particle acceleration.
Such effects produce distinct, observable signatures across the electromagnetic spectrum.

Insight into aspects of the behaviour of flux tubes has been gained through a medley of observational evidence, analytical deduction, and numerical modelling.
A common approach in modelling flux tubes is to construct straight magnetic cylinders linking two parallel photospheric planes \citep[representing the solar surface, following][]{paper:Parker1972}.
Ideal MHD kink-mode instabilities, which presuppose a large degree of twist, result in a \lq\lq kinking\rq\rq\ of the axis and destabilization.
Advances in understanding coronal heating using this hypothesis include those of \citet[][]{paper:BrowningVanderLinden2003, paper:Browningetal2008, paper:Hoodetal2009, paper:Barefordetal2010, paper:Barefordetal2013}.
Such models have enabled forward modelling and comparison with observation \citep{paper:HaynesArber2007, paper:Bothaetal2012, paper:Pintoetal2015, paper:Snowetal2017}, as well as assessing the role played by certain physical effects, or combinations thereof, known to be present in the solar corona \citep[such as by][]{paper:Bothaetal2011, paper:Realeetal2016}.
Straight cylindrical models neglect curvature of flux tubes, which must arc between two photospheric footpoints and whose cross-sectional area may expand from footpoint to apex.
Whether, how, and when instabilities may occur can vary: straight cylinders and the semi-tori representing curved coronal loops are merely topologically, and not geometrically, equivalent.

Of particular interest here is the work of \citet{paper:Barefordetal2016} examining the influence of geometry upon certain properties of a single, highly twisted flux tube.
Comparisons of straight cylindrical flux tubes and curved models suggest that curvature tends to reduce the twist necessary for a kink instability, causing current density to concentrate near the apex (in contrast with a more uniform distribution in straight cylindrical tubes).

While models of single flux tubes have continued to evolve, so too have observational techniques.
One recent focus has been to probe the substructure within the tubular architecture readily apparent in the active corona.
Torsional flows along remarkably twisted atmospheric formations have been reported \citep{paper:Cirtainetal2013,paper:DePontieuetal2014}.
In turn, this has affected modelling, with a recent focus upon loops containing multiple, interacting threads.
Such models have, to date, treated flux tubes between two separate, parallel planes \citep[per][]{paper:Parker1972}.
Examples include threads initialised close to a stability threshold and are capable of destabilising other threads \citep{paper:Tametal2015,paper:Hussainetal2017}, which can thereafter release a great deal of energy through the acceleration of particles \citep[as demonstrated by][]{paper:Threlfalletal2018a}, and which, more importantly, can trigger a runaway \lq\lq avalanche\rq\rq\ process of successive destabilisations in, and the associated releases of energy from, many threads \citep{paper:Hoodetal2016}.
On the other hand, photospheric driving, in addition to causing existing flux tubes to interact and reconnect by relative footpoint motion \citep{paper:OHaraDeMoortel2016}, can also create threads that subsequently interact and mutually destabilise \citep{paper:Reidetal2018,paper:Reidetal2020a}.

The present work aims to explore how curvature affects the process and properties of energetic release, caused by driven photospheric motions in a multi-threaded model, and, in particular, whether (and, if so, how) an MHD avalanche process can take place in such a three-dimensional (3D) MHD model of an arcade, wherein all the threads start and end in the same plane.
Similarly, how far the results of previous models of MHD instabilities and avalanche processes (with straight cylindrical flux tubes) hold for this toroidal geometry, remains to be evaluated.
Key to the avalanche process is that the magnetic field is able to store excess energy and then quickly release it once a threshold is reached. 
The rapid release of magnetic energy occurs when an ideal MHD instability is triggered, and the subsequent restructuring of the field creates more current sheets and an avalanche of releases of energy.

Issues to be addressed include how the field itself can affect the onset, evolution, and energetic output of instabilities among many threads.
In order to address these and other questions, Section~\ref{sec:model} describes our model; Section~\ref{sec:x1} reports the instability of a single thread in a magnetic arcade and Section~\ref{sec:x7i} and Section~\ref{sec:x7m} extend the model, displaying results from the simulation of seven threads subject to different initial driving profiles.
Section~\ref{sec:disc} presents an analysis of these results, before Section~\ref{sec:conc} outlines our conclusions.

\section{Model} \label{sec:model}

The MHD equations are solved in a Lagrangian remap scheme, described by \citet{paper:LareXd2001}.
In non-dimensionalized form, Lare3d solves:
\begin{subequations} \label{eq:MHD}
\begin{eqnarray}
	\frac{\partial{\rho}}{\partial{t}} + \nabla \cdot (\rho\,\vec{v}) &=& 0 \label{eq:continuity} \\
	\rho \frac{\partial{\vec{v}}}{\partial{t}} + \rho \vec{v} \cdot \nabla \vec{v} &=& - \nabla p + \vec{j}\times\vec{B} + \vec{F}_{\mathrm{visc.}} \label{eq:motion} \\
	\frac{\partial{\vec{B}}}{\partial{t}} &=& \nabla \times (\vec{v}\times\vec{B}) + \eta \nabla^2 \vec{B} \label{eq:induction} \\
	\frac{\partial{P}}{\partial{t}} + \vec{v}\cdot \nabla P &=& - \gamma P\,\nabla \cdot \vec{v} + \eta j^2 + q_{\mathrm{visc.}} \label{eq:energy}
\end{eqnarray}
\end{subequations}
for plasma density $\rho$, velocity $\vec{v}$, thermal pressure $P$, magnetic field $\vec{B}$, and associated current density ${\vec{j}}=\nabla\times\vec{B}$.
(Vacuum permeability is $\mu_{0}=4\,\pi\times 10^{-7}\,\mathrm{H}\,\mathrm{m}^{-1}$, which, in dimensionless terms, becomes unity, and $\sigma$ is the usual electrical conductivity.)
Here, the ratio of specific heat capacities is $\gamma=\frac{5}{3}$, and the code ensures divergence-free solutions for the magnetic field [$\nabla\cdot\vec{B}=0$].

The dimensionless variables, as evolved in time in Lare3d, are calculated assuming a magnetic field strength 
$B_{0}$, length $L_0$, and mass density $\rho_{0}$.
Normalising quantities are here chosen as $B_{0}=10^{-3}\,\mathrm{T}=10\,\mathrm{G}$, $L_{0}=10^{7}\,\mathrm{m}=10\,\mathrm{Mm}$, and $\rho_{0}=1.67\times 10^{-12}\,\mathrm{kg}\,\mathrm{m}^{-3}$, leading to the typical and normalising values in Table~\ref{tab:norm}.
Henceforth, lengths are quoted normalised with reference to $L_0$, and times with reference to Alfv\'{e}n times, $\tau_{\mathrm{A}}$.

Our computational domain is $x \in \left[-x_{\mathrm{max.}},x_{\mathrm{max.}}\right], y \in \left[-y_{\mathrm{max.}},y_{\mathrm{max.}}\right], z \in \left[0,z_{\mathrm{max.}}\right]$.
Boundary conditions prescribed in $x$ are periodic (representing neighbouring arcades, which are common in active regions).
Those in $y$ and $z$ are static, perfectly conducting and with zero normal derivatives.
Identical conditions exist on the lower $z$-boundary (our photosphere) for all variables except velocity, in which the driver is imposed according to Section~\ref{ssec:drive}.
Our results and conclusions do not depend on the actual choice of boundary conditions (other than the form of the driver).

\begin{table}[t]
	\caption{Normalising quantities.
	A specific physical quantity listed in the first column, is denoted by the symbol shown in the second, and normalised with reference to the value in the third column.}
	\label{tab:norm}
	\begin{tabular}{lcr@{$\,$}l}
		\hline
		Quantity & Symbol & \multicolumn{2}{c}{Normalising value} \\
		\hline
		magnetic field strength & $B_{0}$ & $1\times 10^{-3}$ & $\mathrm{T}$ \\
		mass density & $\rho_{0}$ & $1.67\times 10^{-12}$ & $\mathrm{kg}\,\mathrm{m}^{-3}$ \\
		length scale & $L_{0}$ & $10\times 10^{6}$ & $\mathrm{m}$ \\
		\hline
		energy density & $W_{0}=\frac{B_{0}^2}{\mu_{0}}$ & $7.96\times 10^{-1}$ & $\mathrm{J}\,\mathrm{m}^{-3}$ \\
		Alfv\'{e}n speed & $v_{\mathrm{A}}=\sqrt{\frac{B_{0}^2}{\mu_{0}\rho_{0}}}$ & $6.90\times 10^{5}$ & $\mathrm{m}\,\mathrm{s}^{-1}$ \\
		Alfv\'{e}n travel time & $\tau_{\mathrm{A}}=\frac{L_{0}}{v_{\mathrm{A}}}$ & $14.5$ & $\mathrm{s}$ \\
		current density & $j_{0}=\frac{B_{0}}{\mu_{0}L_{0}}$ & $7.96\times 10^{-5}$ & $\mathrm{A}$ \\
		magnetic diffusivity & $\eta_{0}=\frac{L_{0}^2}{\tau_{\mathrm{A}}}$ & $6.90\times 10^{12}$ & $\mathrm{m}^{2}\,\mathrm{s}^{-1}$ \\
		\hline
	\end{tabular}
\end{table}

\subsection{Dissipation} \label{ssec:dissipation}

Shock viscosities are used to capture discontinuities and other steep gradients, which otherwise would not be resolved on the numerical grid.
Such viscosities contribute to the viscous dissipation in the system; further details have been given by \citet{thesis:Reid}.
Viscosity enters as a force term [$\vec{F}_\mathrm{visc.}$] in Equation~\ref{eq:motion} and a heating term [$q_\mathrm{visc.}$] in Equation~\ref{eq:energy}.

We apply an anomalous magnetic diffusivity [$\eta_a$]; this is active where some measure of current exceeds a critical threshold.
Such an approach reflects the enhanced resistivity (above classical values) believed to occur in extreme conditions in the corona, such as in very strong current layers.

Slow photospheric motions create each thread.
Over time, these motions also generate internal currents within each thread.
Accordingly, the magnetic energy of the loop system will increase above the initial potential value, and it is this \lq\lq free\rq\rq\ energy that can be tapped when an instability is eventually triggered.
(Theoretically, all energy above that in the potential field is available to be dissipated, but only the energy above that in the linear, force-free, constant-$\alpha$ field of equal helicity is readily accessible.)
At the onset of instability, a thin laminar current sheet rapidly forms along part of the thread; such current is generally faster growing, and ultimately far stronger, than the internal current of the thread.
Typically, in previous investigations \citep[as in that of][]{paper:Tametal2015}, a threshold is chosen based upon the magnitude of current [$\left|{\vec{j}}\right|$].
This threshold would activate anomalous resistivity, having been breached by the current sheet of the instability, but not by internal currents.
Two aspects of the present model warrant a different approach.

Firstly, the curved initial magnetic field, outlined in Section~\ref{ssec:field}, whose strength diminishes with height, naturally gives rise to significant currents near the footpoints of any thread.
Twisting creates a toroidal current; in this geometry, that component is predominantly $j_{z}$ near the boundaries.
Secondly, an additional problem for cases with more than one thread is variation of field strength with height, causing threads closer to the lower boundary to exhibit stronger magnetic field.
Such threads possess, in general, stronger internal currents, compared with outer (higher) threads.
Therefore, the criterion for enhanced resistivity must be triggered by specific current sheets and account for both internal thread currents and variation of $\left|\vec{B}\right|$ with height in the model.
To address these issues, we define a variable:
\begin{equation}
	\zeta = \frac{\sqrt{j_{x}^{2}+j_{y}^{2}}}{\left|\vec{B}\right|},
\end{equation}
and a threshold $\zeta_{\mathrm{crit.}}$.
In this way, magnetic diffusivity $\eta$ {is} $\eta_{\mathrm{a}}=10^{-4}$ where $\zeta\ge\zeta_{\mathrm{crit.}}$, otherwise $\eta=0$. 
This allows strong current sheets related to an instability to be targeted with anomalous resistivity (while ignoring strong currents associated with twisting in low-lying regions of greater field strength).
Although resistivity depends upon $\zeta$ (rather than the more usual $\left|{\vec{j}}\right|$), the effects of resistivity upon the evolution of the magnetic field, in Equation~\ref{eq:induction}, and in the energy equation, Equation~\ref{eq:energy}, remain unchanged.
Ohmic heating, in particular, remains the familiar $\eta j^{2}$ term in the energy equation.

\subsection{Background Magnetic Field} \label{ssec:field}

\begin{figure}[t]
\centering
	\resizebox{0.99\textwidth}{!}{\includegraphics{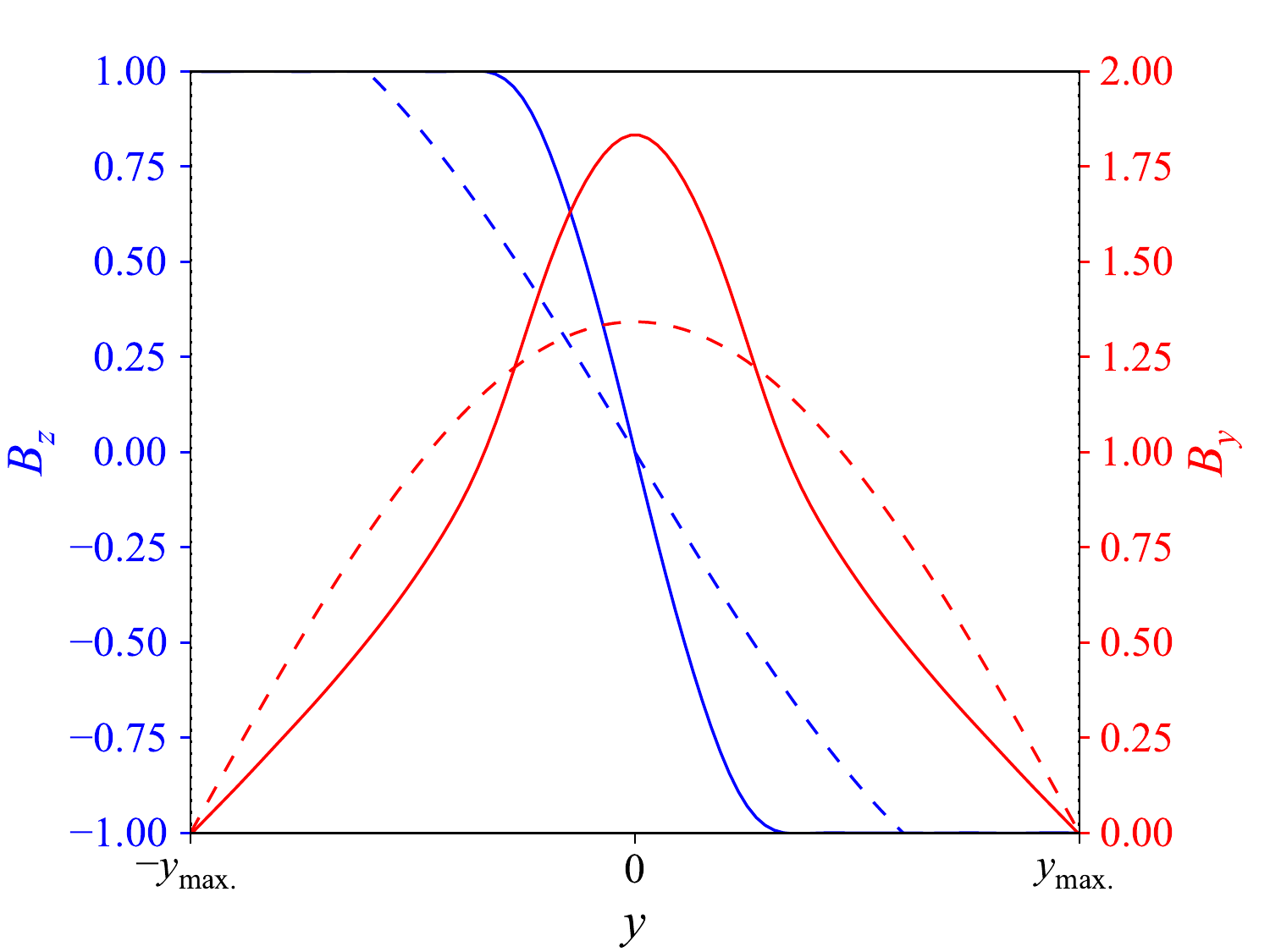}}
	\caption{Components of the modelled magnetic field.
	Vertical ($B_{z}$, seen in blue) and horizontal ($B_{y}$, in red) field components, at $z=0, t=0$, for a model incorporating one Fourier mode or twenty modes in Equation~\ref{eq:B} (shown with dashed or solid lines, respectively); $x$ is the invariant direction in the model.}
	\label{fig:IniB}
\end{figure}

An arcade-like background magnetic field could be constructed in several ways.
In light of our intention to drive the field at the base, two features in particular are desirable: firstly, a vertical photospheric field which reverses sign at the centre of the domain and, secondly, sufficiently large regions of (near-)uniform field strength on either side of a polarity inversion line (PIL).
To that end, we modify a hyperbolic trigonometric field structure \citep[as used by][]{paper:Howsonetal2020}, adding further Fourier modes to create wide regions of near-uniform photospheric $B_{z}$ at the edges of the arcade, while smoothly reversing sign at the domain centre. 
The general form of the field is
\begin{subequations} \label{eq:B}
	\begin{eqnarray}
		B_{y}\left(y,z\right) = \sum_{j=1}^{N}{ a_{j} \frac{\cosh{\left(jk\left(z-z_{\mathrm{max.}}\right)\right)}}{\sinh{\left(-jkz_{\mathrm{max.}}\right)}}\cos{\left(jky\right)} } \label{eq:By} \\
		B_{z}\left(y,z\right) = \sum_{j=1}^{N}{ a_{j} \frac{\sinh{\left(jk\left(z-z_{\mathrm{max.}}\right)\right)}}{\sinh{\left(-jkz_{\mathrm{max.}}\right)}}\sin{\left(jky\right)} }, \label{eq:Bz}
	\end{eqnarray}
\end{subequations}
in which $k=\frac{\pi}{2y_{\mathrm{max.}}}$.
Coefficients $a_{j}$ are constructed for the several Fourier modes as:
\[
	a_{j} = \left\{
		\begin{array}{cc}
			\frac{3\,\pi+8}{18\,\pi} & j=3 \\
			\frac{1}{9\,\pi} & j=6 \\
			18\frac{\left(64c^6+32c^5-96c^4-40c^3+36c^2-2\right)\left(j^2-9\right)+9c\left(j^2-6\right)}{j\pi\left(j^2-36\right)\left(j^2-9\right)} & j\ne3,6
		\end{array}
	\right., \nonumber
\]
with $c=\cos{\left(\frac{j\pi}{6}\right)}$.
Preliminary tests suggested that fixing $N=20$ yields a sufficiently wide region of approximately uniform magnetic field strength, which could contain several threads.
Figure~\ref{fig:IniB} illustrates the magnetic field profile seen in Equation~\ref{eq:B}, comparing cases where $N=1$ and $N=20$.
The size of the regions of uniform magnetic field strength also increases with the number of Fourier modes.

In tests, simulations with $N=1$ resulted in excessively strong currents at the footpoints of each thread in the domain, which grew rapidly over time in response to an imposed rotation.
Variation of magnetic field strength over the radius was responsible; these difficulties are avoided if $N\geq 20$.

\subsection{Photospheric Driver} \label{ssec:drive}

Within the magnetic arcade, a vortical driving motion is applied at the footpoints of the thread(s) to be created, twisting the magnetic field.
In order to aid comparison with previous cases of photospherically driven flux tubes, the form of the driver matches that imposed by \citet[][]{paper:Reidetal2018}:
\begin{subequations} \label{eq:Vphi}
	\begin{eqnarray}
		v_{\phi}\left(r,t\right) &=& v_{0}f\left(r\right)D\left(t\right) \label{eq:Vphi_Form} \\
		f\left(r\right) &=&\left\{
			\begin{array}{ll} 
				\frac{r}{a}\left(1-\frac{r^2}{a^2}\right)^3 & \qquad\qquad\qquad r\le a \\
				0 & \qquad\qquad\qquad r>a
			\end{array}
		\right. \label{eq:Vphi_Fr} \\
		D\left(t\right) &=&\left\{
			\begin{array}{ll} 
				0 & \qquad t \leq t_{\mathrm{s}} \\
				\frac{1}{2}\left(1-\cos{\left(\frac{(\pi\left(t-t_{\mathrm{s}}\right)}{t_{\mathrm{e}}-t_{\mathrm{s}}}\right)}\right) & \qquad t_{\mathrm{s}}< t<t_{\mathrm{e}} \\
				1 & \qquad t \geq t_{\mathrm{e}}.
			\end{array} \right. \label{eq:Vphi_Dt}
	\end{eqnarray}
\end{subequations}
Driving is imposed over a radius $a$, the minor toroidal radius of the flux tube.
Velocity, with amplitude governed by $v_{0}$, is gradually introduced through a function $D\left(t\right)$.
In the straight cylindrical case, the spatial form of the driver produced zero net axial current.
However, the fact that the axial field here is not 
completely uniform at the photospheric boundary means that the net current is small but non-zero.
While \citet{paper:Reidetal2018} use a driver beginning at $t=0$, we postpone the start of rotation in order that the magnetic arcade relaxes towards potential.
The ramp-up phase begins at time $t_{\mathrm{s}}$ and ends at $t_{\mathrm{e}}$, after which $v_{\phi}\left(r,t\right)$ remains constant.
Initial tests have determined that a delay of $50\,\tau_{\mathrm{A}}$ allows the model to reach sufficiently close to minimal energy (hence $t_s=50\,\tau_{\mathrm{A}}$), and that a duration extent of $10\,\tau_{\mathrm{A}}$ is sufficient to avoid unnecessary shocks (and hence $t_e=60\,\tau_{\mathrm{A}}$).

In Section~\ref{ssec:field}, we noted that our model aimed to fit many threads in the near-uniform field region; the size of this region places constraints on the radius $a$.
The radius must be large enough to resolve the formation of associated current sheets, while being small enough to fit many threads in the near-uniform field region.
Similarly, our choice of velocity amplitude $v_{0}$ is also constrained by the need to be faster than slow numerical diffusion at the base, yet slower than the coronal Alfv\'{e}n speed.
This issue of timescale separation is addressed by \citet{paper:Bownessetal2013}.

Previous investigations have shown that the relative amplitude of the photospheric driver in different threads plays a key role in determining the sequence in which straight cylindrical multi-threaded flux tubes destabilise \citep{paper:Reidetal2018}.
We aim to extend this to consider a set of a toroidal loops, and to study the impact of threads of different length, curvature, and driving speed.
\begin{figure*}[t]
	\centering
	\subfloat[Initial magnetic structure.]{\label{fig:Initial_x1FL}\includegraphics[width=0.53\linewidth, clip=true,  trim=35 45 85 105]{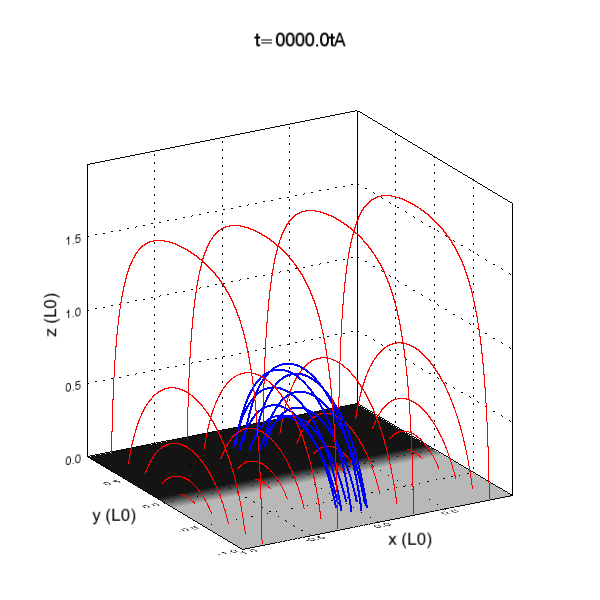}}%
	\subfloat[Velocity $v_{x}, v_{y}$.]{\label{fig:Initial_x1VelocityPlane}\includegraphics[width=0.46\linewidth, clip=true, trim=20 75 55 0]{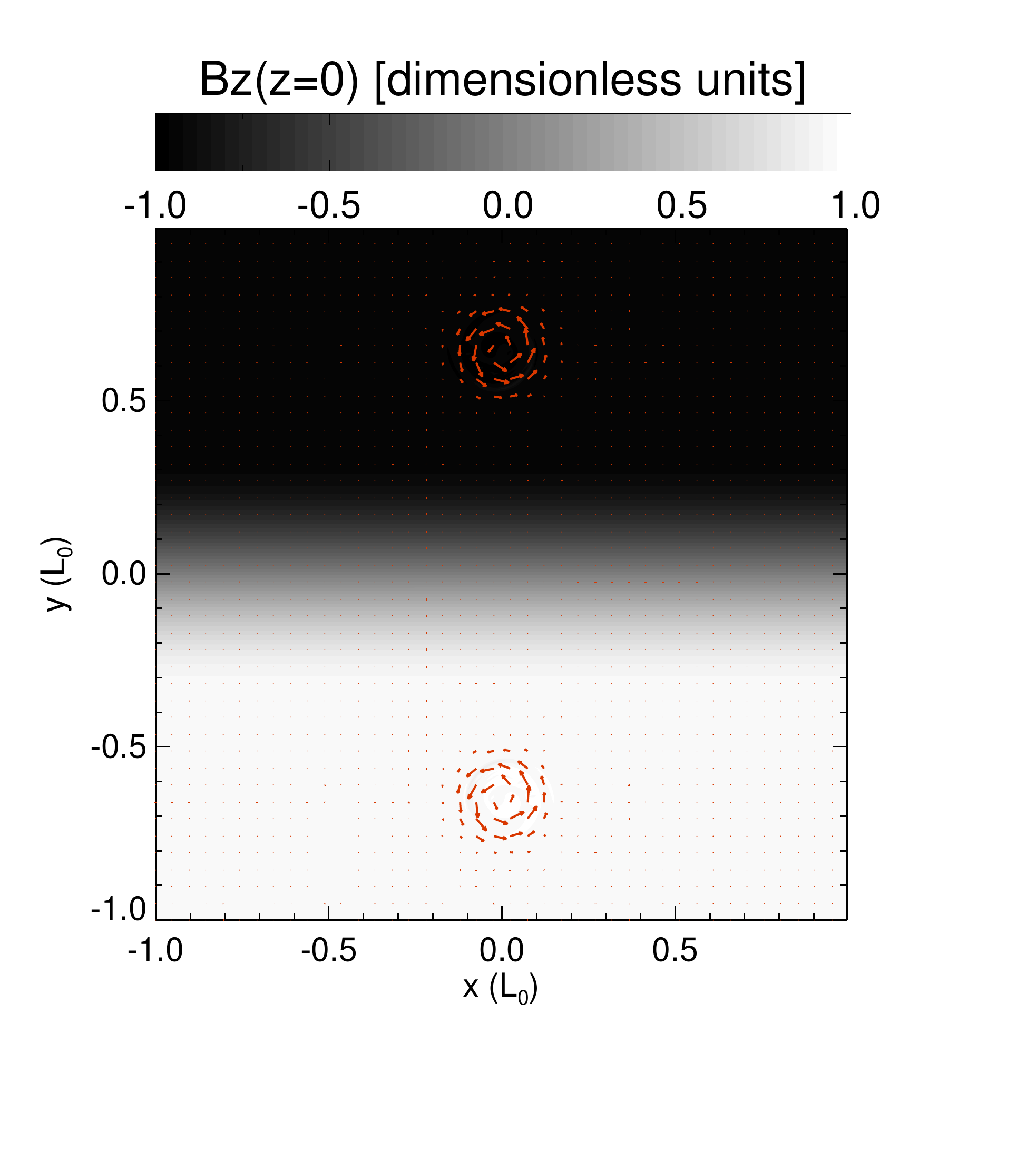}}
	\caption{Single thread: field and driver. \protect\subref{fig:Initial_x1FL} Field structure at $t=0$, showing field lines traced throughout the domain (red) and specifically chosen field lines traced from the driven region (blue), above a magnetogram showing vertical field strength at the base ($B_z\left(z=0\right)$).
	\protect\subref{fig:Initial_x1VelocityPlane} Contours of the local vertical field strength at the base of the simulation, with planar velocity vectors ($v_{x}$, $v_{y}$) overlaid, illustrating photospheric driving at the base at full speed ($t=t_e$).}
	\label{fig:Initial}
\end{figure*}

\subsection{Instability}\label{subsec:inst}

The onset of an MHD avalanche, in a flux tube containing many threads, requires a trigger.
An exponential rise in the kinetic energy accompanies the first disruption (seen later in, for example, Figure~\ref{fig:x1En}). This rise is due to an ideal MHD instability, which begins the avalanche process.
It is possible that the ideal kink instability, widely used in straight cylindrical cases \citep[e.g.][]{paper:Tametal2015,paper:Hussainetal2017}, is responsible.
A sufficiently twisted thread will \lq\lq kink\rq\rq\ (i.e. displace radially outward), with this mode characterised by the rapid development of a helical current sheet surrounding the unstable thread.
The twist [$\Phi$] along field lines necessary for this instability is a critical parameter.
However, critical thresholds \citep[and, indeed, how one quantifies twist, discussed, for example, by][]{paper:Threlfalletal2018b,paper:Threlfalletal2020} are subject to considerable uncertainty, and may deviate significantly from theoretical values, which are usually specific to particular field configurations; for example, $\Phi\gtrapprox 3.3\,\pi$ for uniformly twisted tubes \citep{paper:HoodPriest1979b}.
For a loop with radially varying speed, \citet{paper:Gerrardetal2004} determine an average twist after a given time [$t$].
When applied to our driving velocity, the same approach gives the average twist at time $t$ here:
\begin{equation}
		\langle\Phi\rangle =\frac{v_{0}}{2a}\left[t-\frac{1}{2}\left(t_{\mathrm{s}}+t_{\mathrm{e}}\right)\right], 
		\label{eq:avetwist}
\end{equation}
(for $t>t_{\mathrm{e}}$; for our profile, twist peaks on the axis at four times this value).

Our introduction of toroidal geometry into our multi-threaded arcade makes the torus instability \citep[][]{book:Bateman1978} a plausible trigger mechanism for the avalanche, although our flux tubes do not establish significant non-zero net current in each thread.
The torus instability is analogous to the kink instability, but it occurs in a different set of conditions, including a large and negative rate of change of magnetic field along the major toroidal axis [$R$].
Such a rate of change is commonly quantified by the decay index $n=-{\mathrm{d}\log{\left|\vec{B}_{\mathrm{ex.}}\right|}}/{\mathrm{d}\log{R}}$, where $\vec{B}_{\mathrm{ex.}}$ denotes the external magnetic field.
{A} field is unstable to the torus instability where $n>n_{\mathrm{crit.}}$.
The critical threshold $n_\mathrm{crit.}$ varies depending on the configuration.
For a purely poloidal external field, \citet[][]{book:Bateman1978} derives $n_{\mathrm{crit.}} = 1.5$, but several effects, such as expansion, can lower such values, or 
raise them as high as $n_\mathrm{crit.}=2$ \citep{paper:KliemTorok2006,paper:TitovDemoulin1999,paper:Aulanieretal2010,paper:Zuccarelloetal2015}.
However, the original derivation of the torus instability featured a non-zero net toroidal current, surrounded by a purely poloidal magnetic field. 
This is not the case here and the condition for instability will be modified.
Thus, the value and importance of $n$ is subject to considerable uncertainty, and so we consider a plausible range of critical $n$ \citep[after][ who also consider a flux tube with zero total axial current in the initial state]{paper:Syntelisetal2017}.

We will provide diagnostic information (i.e. the twist and decay index at each thread destabilisation) for additional context.
While it is interesting to determine if the critical conditions at the onset of the ideal MHD instability correspond to either a kink or torus mode, the important point, for the avalanche, is that there is \emph{an} ideal MHD instability. This instability must grow on an Alfv\'{e}n timescale.

\section{Single-Threaded Case} \label{sec:x1}

\begin{figure*}
	\includegraphics[keepaspectratio, width=\linewidth]{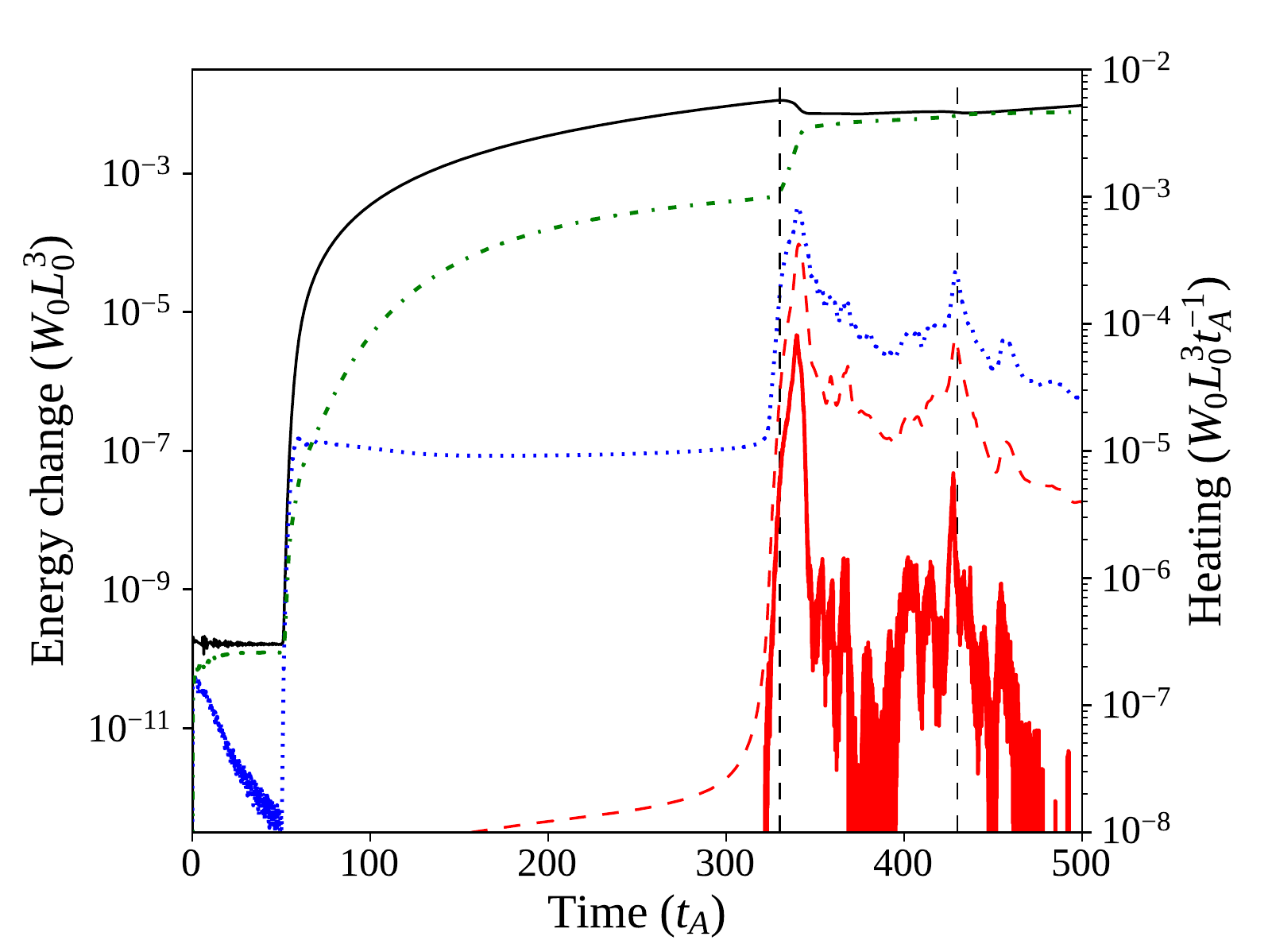}%
	\caption{Single-threaded case, energies: Magnetic (solid, black), internal (dash-dotted, green), and kinetic (dotted, blue) energy components with their initial values subtracted, shown together with Ohmic (thick, red) and viscous (dashed, red) instantaneous heating in the system.
	The two dashed, vertical lines indicate times at which strong current sheets form, illustrated in contours of current in Figure~\ref{fig:x1J}, and hence there are large releases of energy.}
	\label{fig:x1En}
\end{figure*}

Before considering the behaviour of a loop containing many threads in the arcade field, described in Section~\ref{ssec:field}, we put this in context by first considering the behaviour of a single, continuously driven thread.
Within a domain with $x_\mathrm{max.}=y_{\mathrm{max.}}=1.0,z_{\mathrm{max.}}=2.0$, the prescribed form of velocity, Equation~\ref{eq:Vphi}, is implemented on both sides of the PIL (found at $y=0$).
The velocity profile is centred on $\left(x,y\right)=\left(0,\pm 0.65\right)$, with a radius $a=0.2$ and with $v_{0}=0.008$.
The general 3D structure of the system can be seen in Figure~\ref{fig:Initial}.
Blue field lines in Figure~\ref{fig:Initial_x1FL} highlight the part of the domain that will be continuously driven, subject to the driving profile seen in Figure~\ref{fig:Initial_x1VelocityPlane}.
The driven magnetic field region becomes increasingly twisted as time progresses, until such time as an ideal instability forms a current sheet in the flux tube.
The volume-integrated magnetic, kinetic, and internal energies and the instantaneous components of heating are seen in Figure~\ref{fig:x1En}.
As the bottom boundary is constantly driven, Poynting flux is continuously being injected into the system.
After the avalanche process has started, the accumulated heating (consisting of Ohmic and viscous heating) eventually settles down to around $35\,\%$ of the time-integrated injected Poynting flux.

Components of energy and heating neatly describe distinct phases of system behaviour.
Prior to photospheric driving (at $t=50\,\tau_{\mathrm{A}}$), there is a slight decrease in magnetic energy and a sharper reduction in kinetic energy as the system relaxes towards minimum energy.
At the onset of driving, all energies increase sharply; while kinetic energy appears to reach a quasi-steady state faster, magnetic and internal energies continue to increase still further, and then gradually begin levelling off.
Viscous heating also follows the start of driving, but it remains very small.
The magnetic field is gradually twisted, while internal currents associated with the thread form (as seen in Figure~\ref{fig:x1J_315}).
\begin{figure*}
	\centering
	\subfloat[$t=330\,\tau_{\mathrm{A}}$]{\label{fig:x1FL_First}\includegraphics[width=0.5\linewidth, clip=true, trim=40 45 85 105]{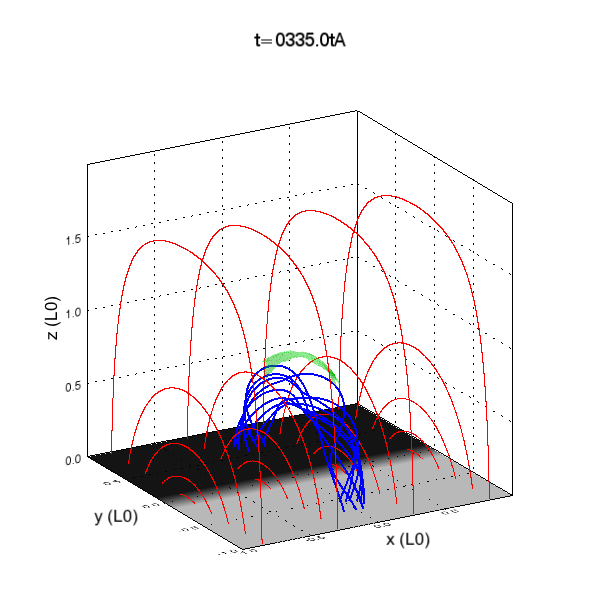}}%
	\subfloat[$t=430\,\tau_{\mathrm{A}}$]{\label{fig:x1FL_Second}\includegraphics[width=0.5\linewidth, clip=true,  trim=40 45 85 105]{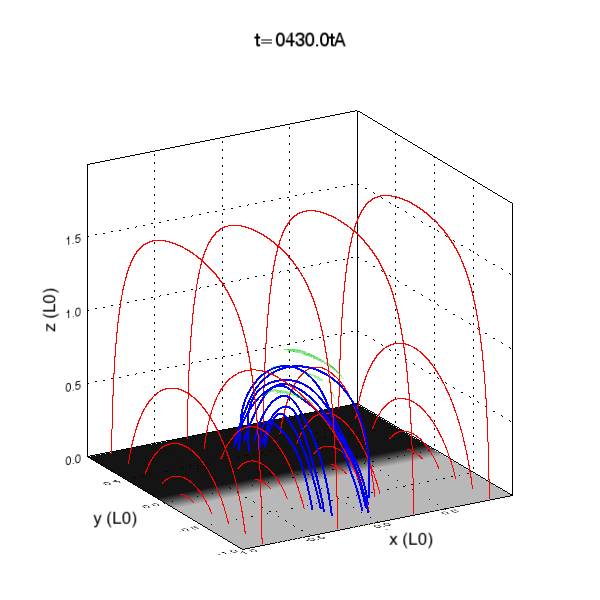}}
	\caption{Single-threaded case, 3D configuration: \protect\subref{fig:x1FL_First} illustrates the 3D configuration of the magnetic field during the initial energy release event (at $t=330\,\tau_{\mathrm{A}}$), while \protect\subref{fig:x1FL_Second} illustrates a secondary (less energetic) eruption ($t=430\,\tau_{\mathrm{A}}$).
Ambient field is shown in red, while field lines traced from the photospheric driving region at the base are shown in blue, with green isosurfaces of current where $\zeta\geq 0.75\,\zeta_{\mathrm{crit.}}$ (lowered below $\zeta_{\mathrm{crit.}}$ for illustrative purposes).}
	\label{fig:x1FL}
\end{figure*}

The second phase of the experiment begins when a current sheet forms along the twisted thread at approximately $t=330\,\tau_{\mathrm{A}}$.
Current in the sheet eventually satisfies the condition $\zeta>\zeta_{\mathrm{crit.}}$, triggering anomalous resistive effects.
An example of the 3D structure of such a current sheet can be seen in Figure~\ref{fig:x1FL_First}; locations which satisfy $\zeta>\zeta_{\mathrm{crit.}}$ are difficult to see in contour images (Figure~\ref{fig:x1J}).
Resistivity, where activated, causes Ohmic heating.
\begin{figure*}
	\centering
	\subfloat[$t=315\,\tau_{\mathrm{A}}$.]{\label{fig:x1J_315}\includegraphics[keepaspectratio, width=0.305\linewidth, clip=true, trim=10 5 10 20]{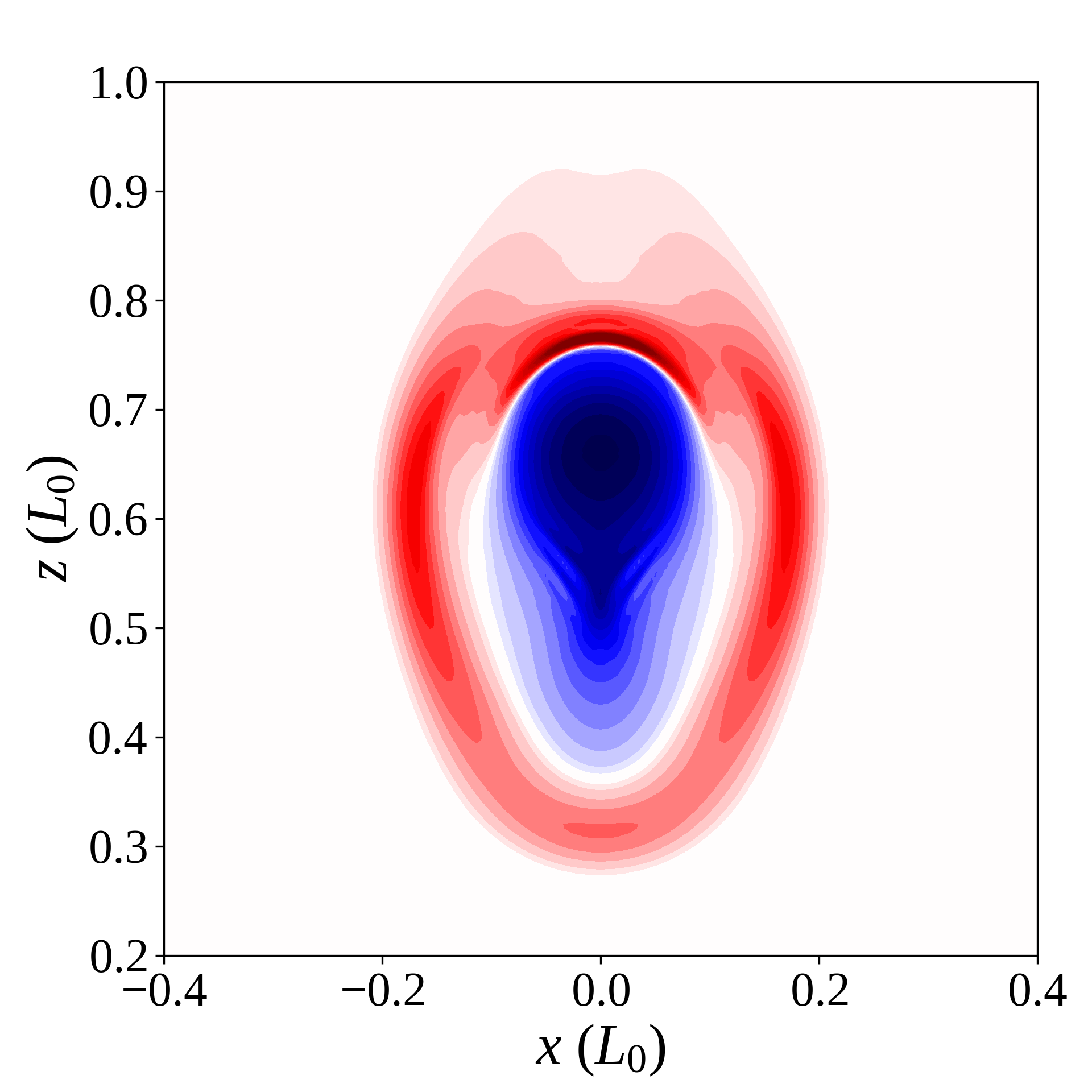}}%
	\subfloat[$t=330\,\tau_{\mathrm{A}}$.]{\label{fig:x1J_330}\includegraphics[keepaspectratio, width=0.305\linewidth, clip=true, trim=10 5 10 20]{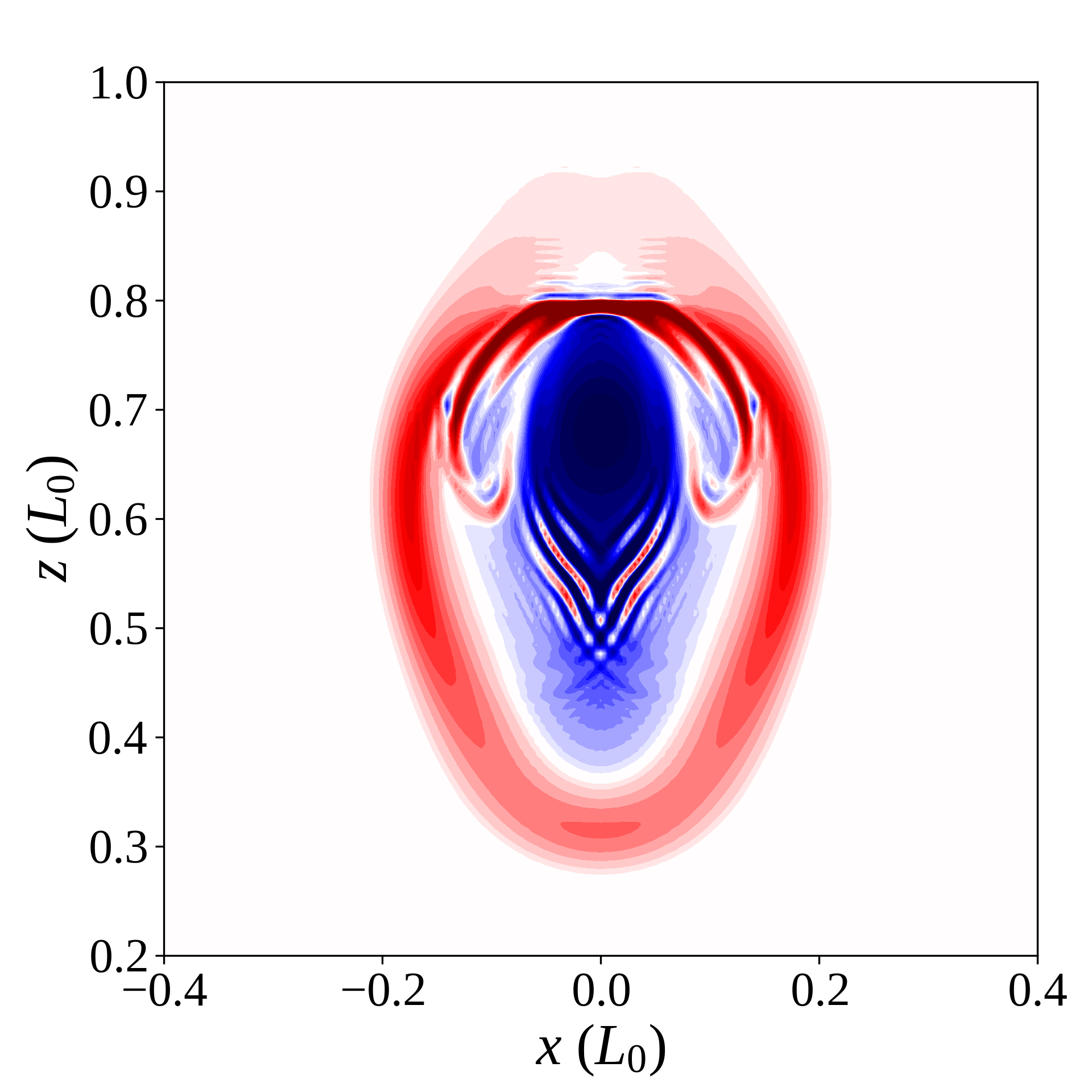}}%
	\subfloat[$t=430\,\tau_{\mathrm{A}}$.]{\label{fig:x1J_500}\includegraphics[keepaspectratio, width=0.305\linewidth, clip=true, trim=10 5 10 20]{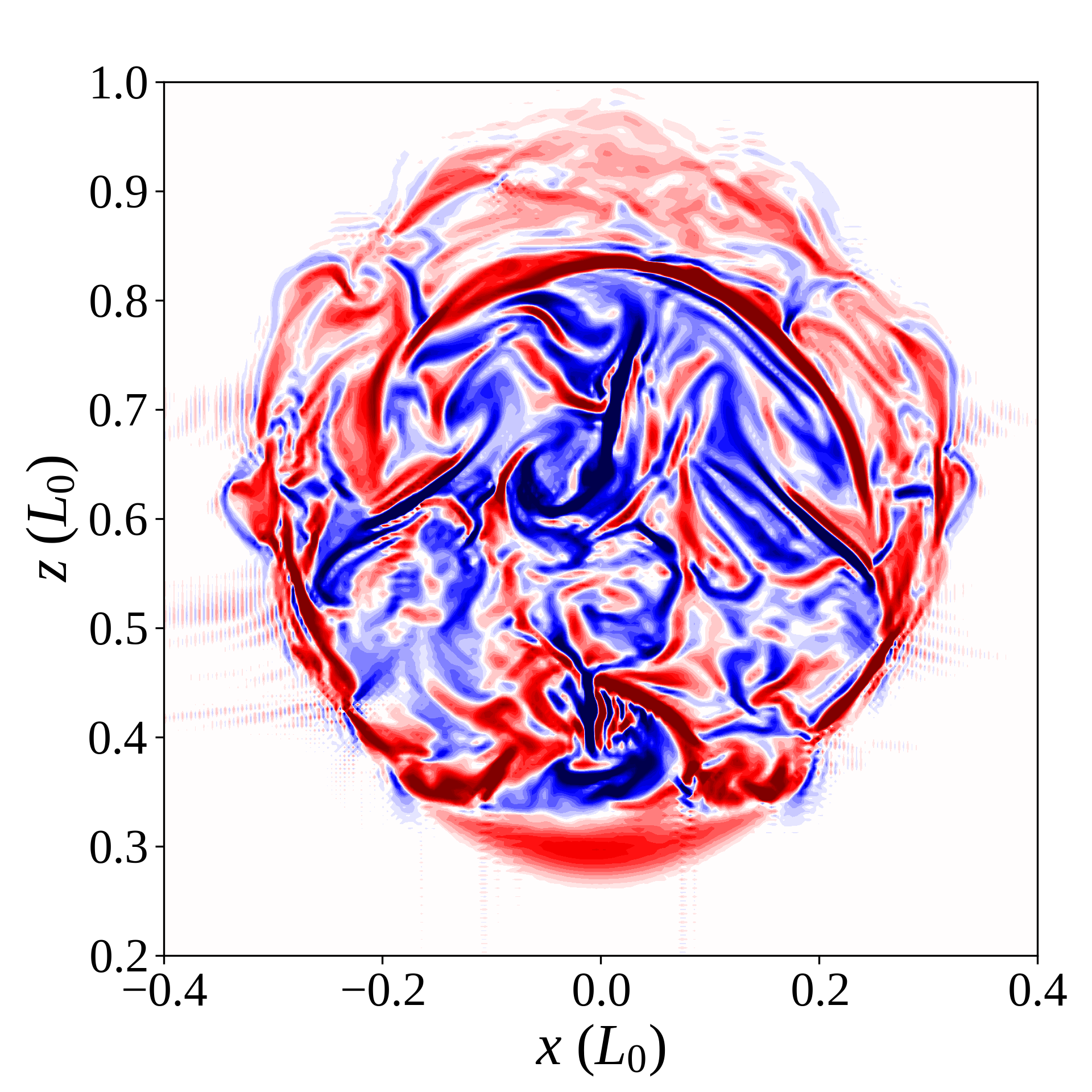}}%
	\includegraphics[keepaspectratio, width=0.085\linewidth, clip=true, trim=13 15 13 15]{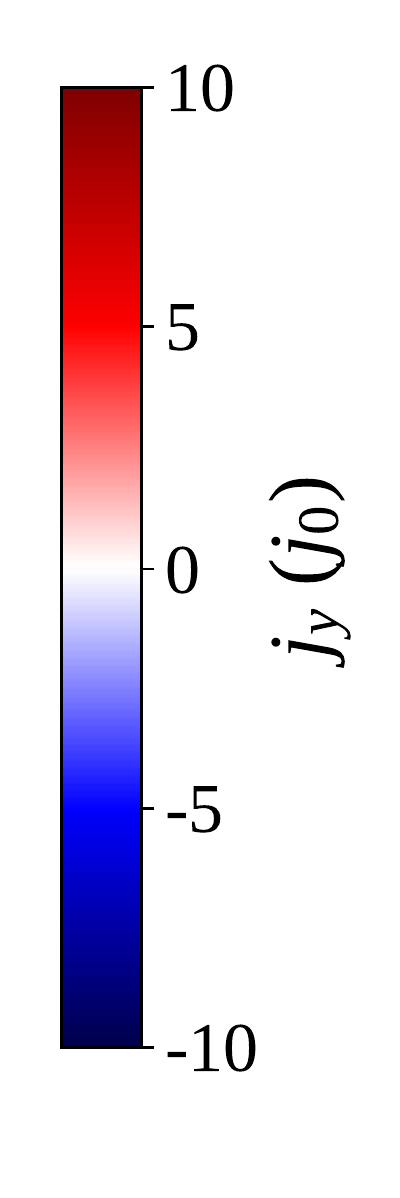}
	\caption{Contours of toroidal current [$j_y$] above the polarity inversion line [$y=0$], in a single-threaded loop.}
	\label{fig:x1J}
\end{figure*}

Following this event, the final phase of behaviour begins, as the tube remnants undergo restructuring and gradual expansion, while the magnetic field continues to be driven at the base.
This phase is characterised by additional, aperiodic formation of small, secondary current sheets, scattered throughout the tube remnants.
A Fast Fourier Transform (FFT) of the energy and heating components (not shown) detects no distinct frequency or period, hence our description of the formation of these current sheets as \lq\lq aperiodic\rq\rq.
Examples of such current sheets are present at $t=430\,\tau_{\mathrm{A}}$, and are visible in Figure~\ref{fig:x1FL_Second} as thin 3D isosurfaces, or as highly concentrated regions of current in Figure~\ref{fig:x1J_500} throughout the loop.
The heating from such current sheets is weaker than, but comparable to, the first.
We end the numerical experiment at $t=500\,\tau_{\mathrm{A}}$, however, there is nothing to suggest that this final phase (of aperiodic formation and dissipation of current sheets) would not continue indefinitely.

Regarding the onset of the initial instability, the thread has an average twist of $1.75\,\pi$ (and maximum $7\,\pi$) at $t=330\,\tau_{\mathrm{A}}$.
Similarly, we recover a decay index $n\approx1.57$ on the axis at the apex; some variation occurs with radius, but $n$ is around, or just below, $1.5$ across most of the cross-section.

During the first major instability, {the flux tube is more twisted} than in later disruptions (Figure~\ref{fig:x1FL_Second}).
Much of the initial twist dissipates in the first reconnection event: the field is more relaxed, and heating arises from the continual conversion of injected {magnetic} energy to thermal/internal energy.
The Poynting flux injected through the driven boundary dissipates at a largely constant rate.
No steady state in energy is achieved: energy releases are aperiodic and vary in size and location within the loop remnants.
Later energy releases are far smaller than, but similar to, the first.
These aspects are readily apparent in the energies displayed in Figure~\ref{fig:x1En}.

\section{Seven Threads: Identical Drivers} \label{sec:x7i}

Our investigation of a single thread puts into context the behaviour observed when additional threads are included in our arcade model.
We now consider seven threads, all created by photospheric driving as before.
The footpoints are hexagonally packed, with a row of three threads between two rows of two threads; the general arrangement can be seen accompanying our later discussion (in Figure~\ref{fig:x7iCNX_0}).
The central thread is centred on $\left(x,y\right)=\left(0,\pm 1.3\right)$.
To accommodate these additional threads, the length of the simulation domain is doubled in $y$ (i.e. $y_{\mathrm{max.}}=2, y\in\left[-2,2\right]$).
All thread footpoints are rotated with $v_{0}=0.02$ over a radius $a=0.1$ (i.e. half that used in the single-threaded case).
\begin{figure*}[t]
	\centering
	\subfloat[Initial state ($t=0$).]{\label{fig:x7iFLa}\includegraphics[width=0.5\linewidth, clip=true,  trim=40 45 85 105]{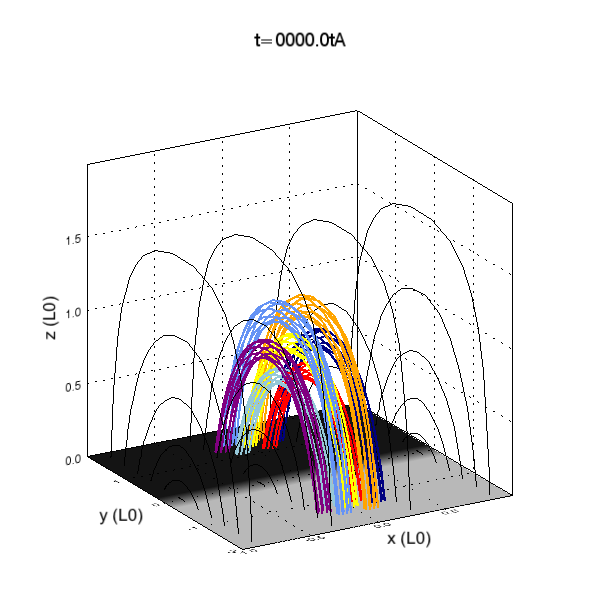}}%
	\subfloat[First instability ($t=150\,\tau_{\mathrm{A}}$).]{\label{fig:x7iFLb}\includegraphics[width=0.5\linewidth, clip=true,  trim=40 45 85 105]{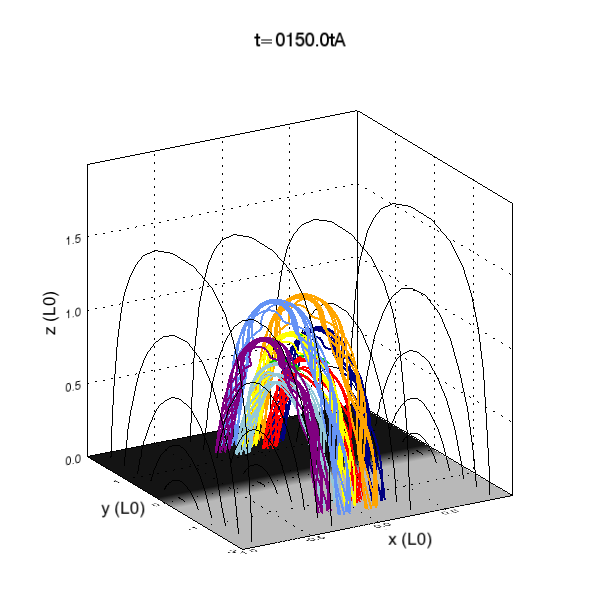}}\\
	\subfloat[Top row destabilises ($t=165\,\tau_{\mathrm{A}}$).]{\label{fig:x7iFLc}\includegraphics[width=0.5\linewidth, clip=true,  trim=40 45 85 105]{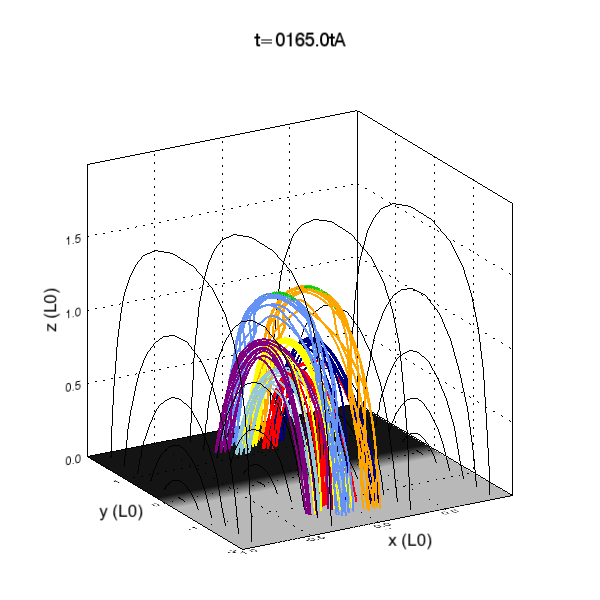}}
	\subfloat[Final state ($t=500\,\tau_{\mathrm{A}}$).]{\label{fig:x7iFLd}\includegraphics[width=0.5\linewidth, clip=true,  trim=40 45 85 105]{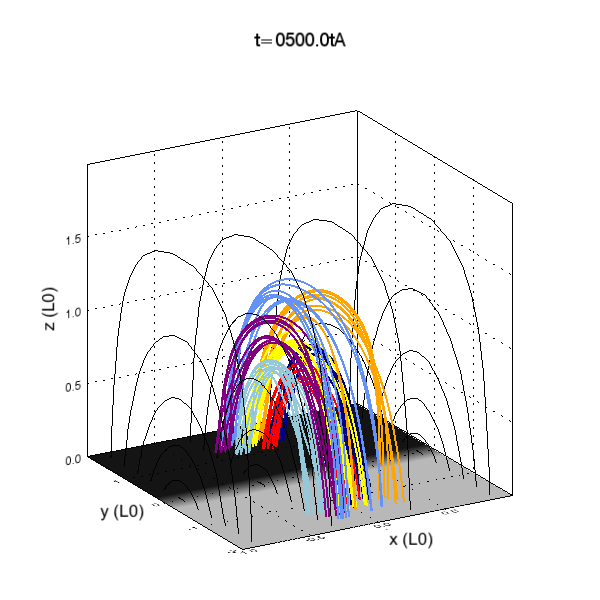}}
	\caption{Seven threads, identical drivers: 3D magnetic configuration at different times during the experiment, where specific field lines have been coloured corresponding to their initial position in one of seven regions of photospheric driving, and green isosurfaces (if present) identify regions where $\zeta>0.75\,\zeta_{\mathrm{crit.}}$ (lowered from $\zeta_{\mathrm{crit.}}$ for illustrative purposes).
For key to colours and destabilisation times, see Table~\ref{tab:threads}.}
	\label{fig:x7iFL}
\end{figure*}

\begin{figure*}[t]
	\includegraphics[keepaspectratio, width=\linewidth]{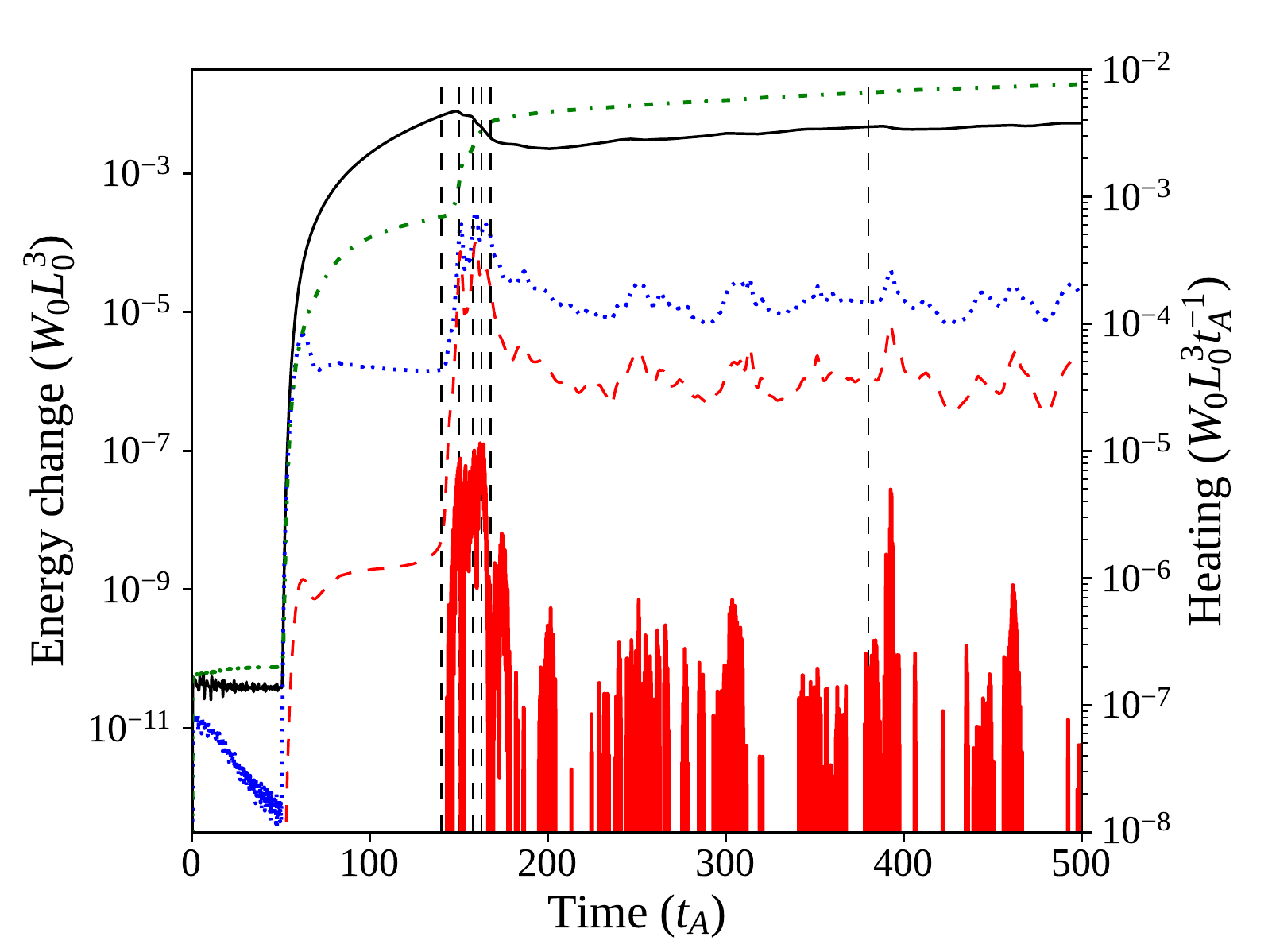}
	\caption{Seven threads, with identical driving speeds, energies: Magnetic (solid, black), internal (dash-dotted, green), and kinetic (dotted, blue) energy components with their initial values subtracted, shown together with Ohmic (thick, red) and viscous (dashed, red) instantaneous heating of the system.
	The six dashed, vertical lines indicate times associated with mergers of threads, illustrated as contours of current in Figure~\ref{fig:x7iJ}.}
	\label{fig:x7iEn}
\end{figure*}

\begin{figure*}[t]
	\centering
	\subfloat[$t=140\,\tau_{\mathrm{A}}$.]{\label{fig:x7iJ_140}\includegraphics[keepaspectratio, width=0.305\linewidth, clip=true, trim=10 5 10 20]{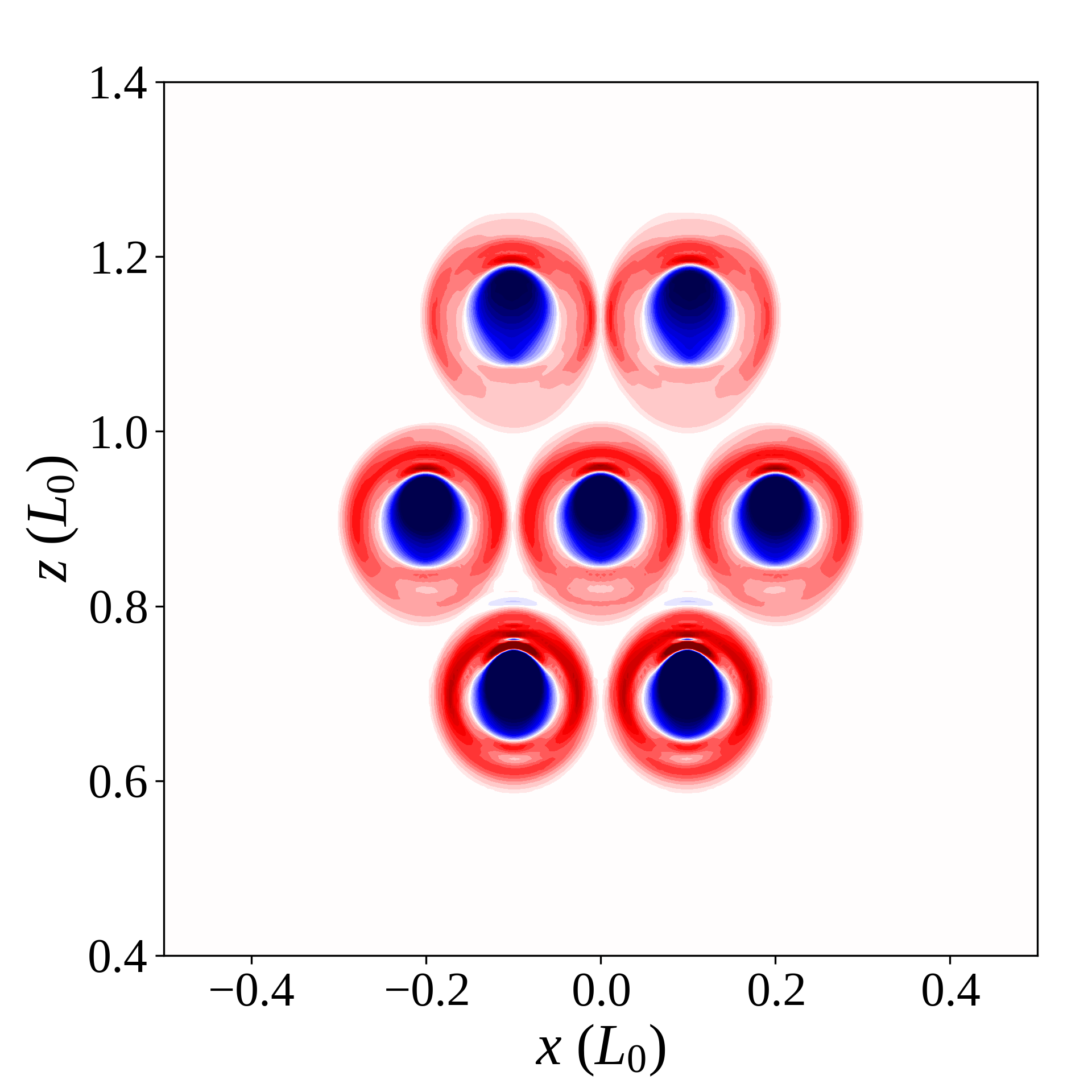}}%
	\subfloat[$t=150\,\tau_{\mathrm{A}}$.]{\label{fig:x7iJ_150}\includegraphics[keepaspectratio, width=0.305\linewidth, clip=true, trim=10 5 10 20]{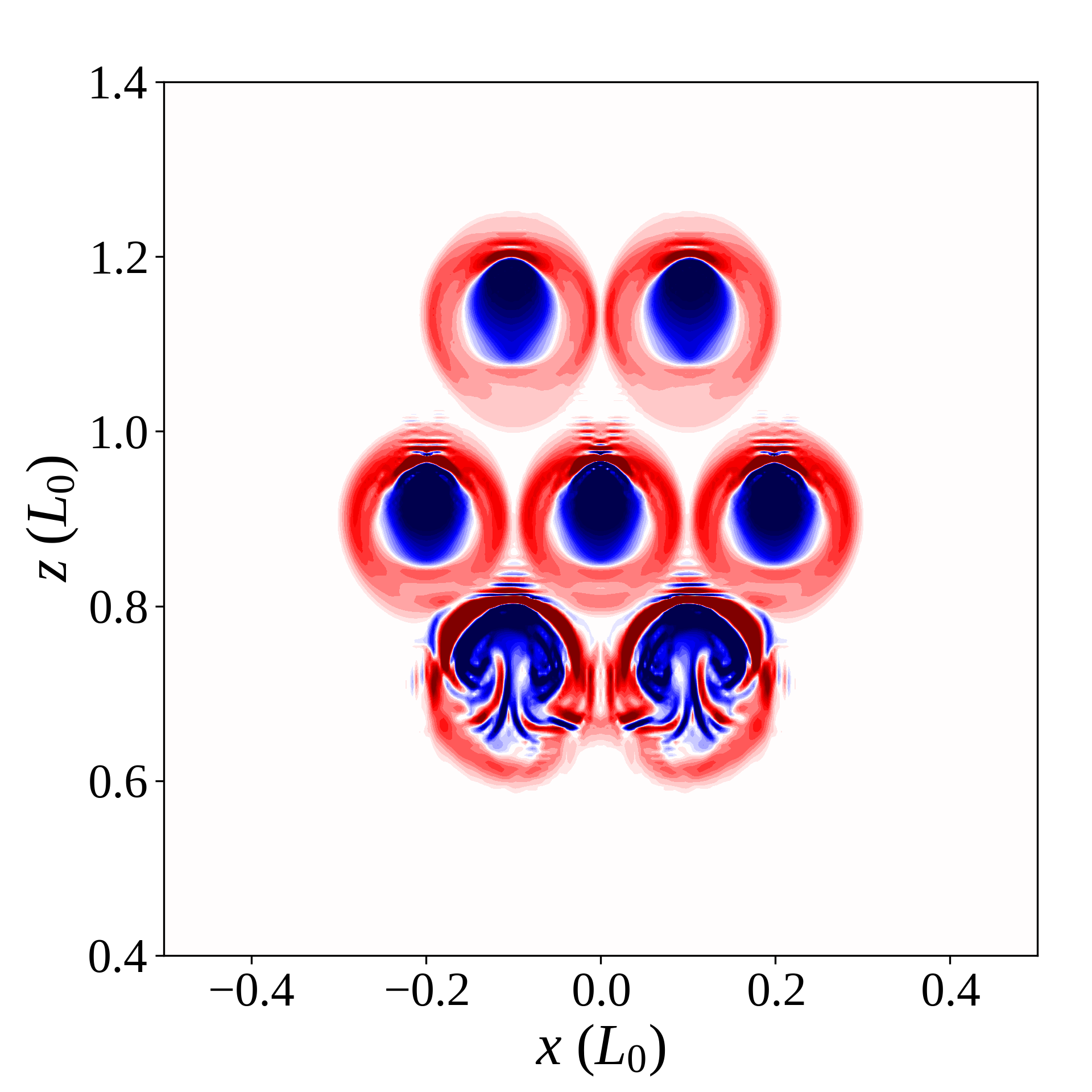}}%
	\subfloat[$t=157.5\,\tau_{\mathrm{A}}$.]{\label{fig:x7iJ_157}\includegraphics[keepaspectratio, width=0.305\linewidth, clip=true, trim=10 5 10 20]{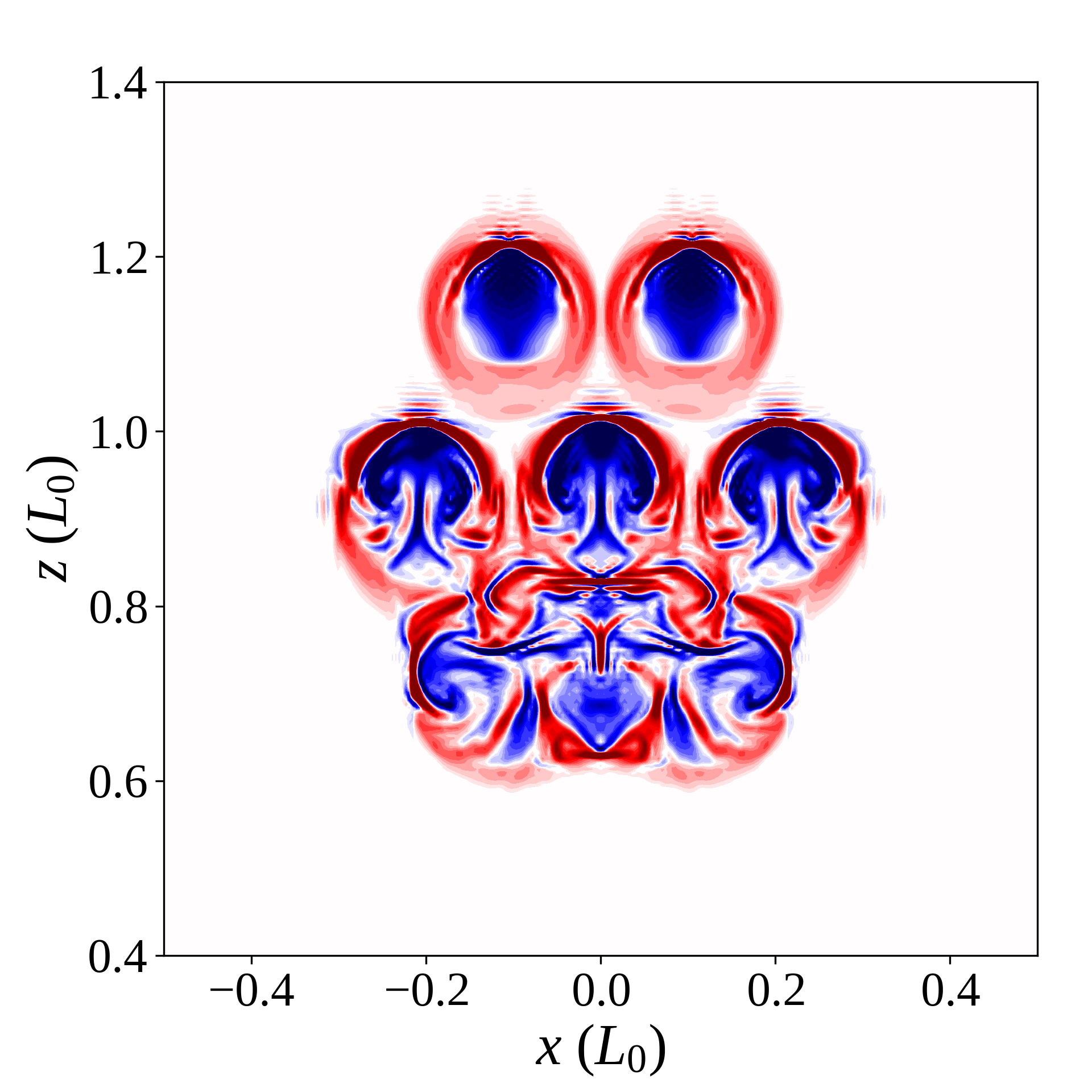}}%
	\includegraphics[keepaspectratio, width=0.085\linewidth, clip=true, trim=13 15 13 15]{JyContour_ColourBar.pdf}\\%
	\subfloat[$t=162.5\,\tau_{\mathrm{A}}$.]{\label{fig:x7iJ_162}\includegraphics[keepaspectratio, width=0.305\linewidth, clip=true, trim=10 5 10 20]{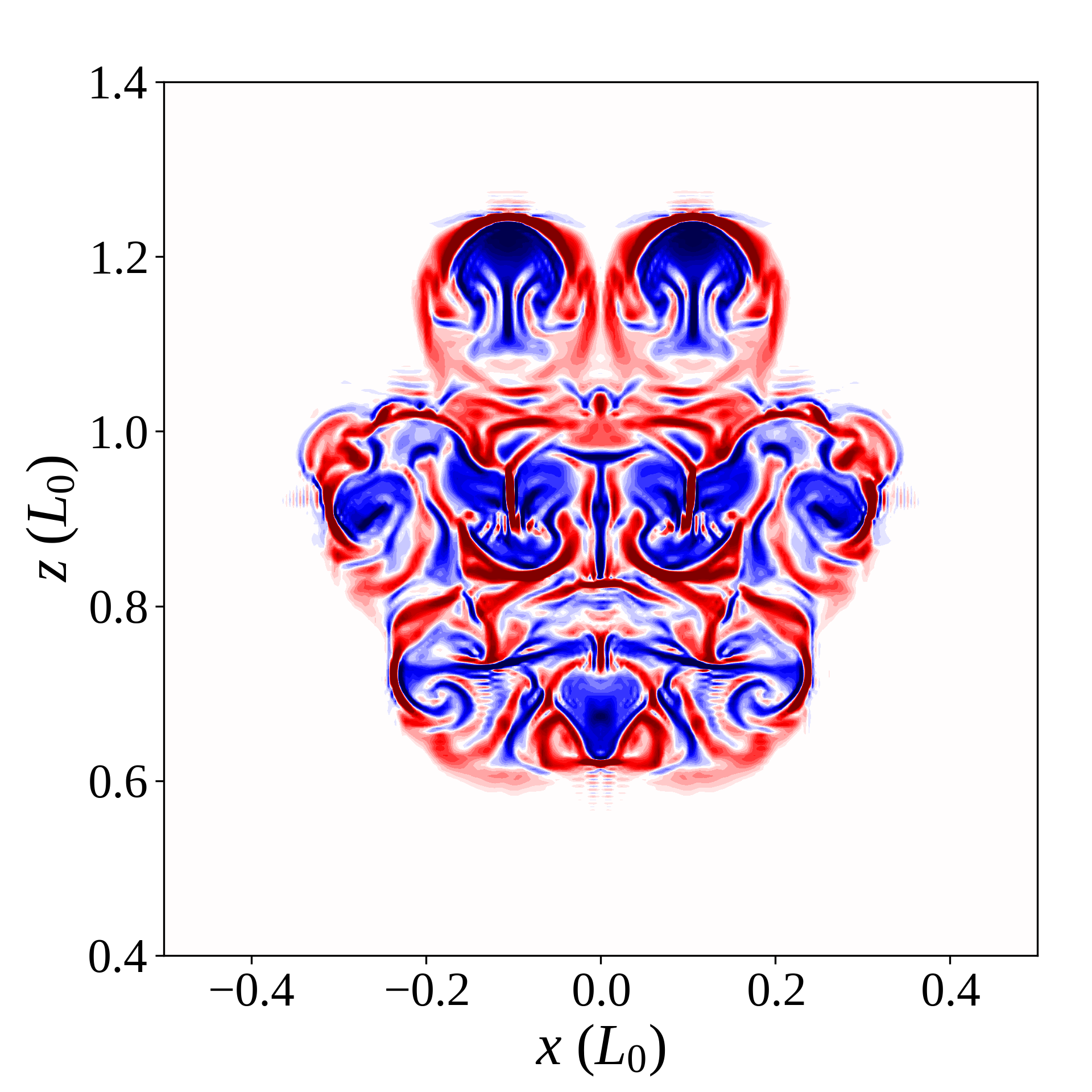}}
	\subfloat[$t=167.5\,\tau_{\mathrm{A}}$.]{\label{fig:x7iJ_167}\includegraphics[keepaspectratio, width=0.305\linewidth, clip=true, trim=10 5 10 20]{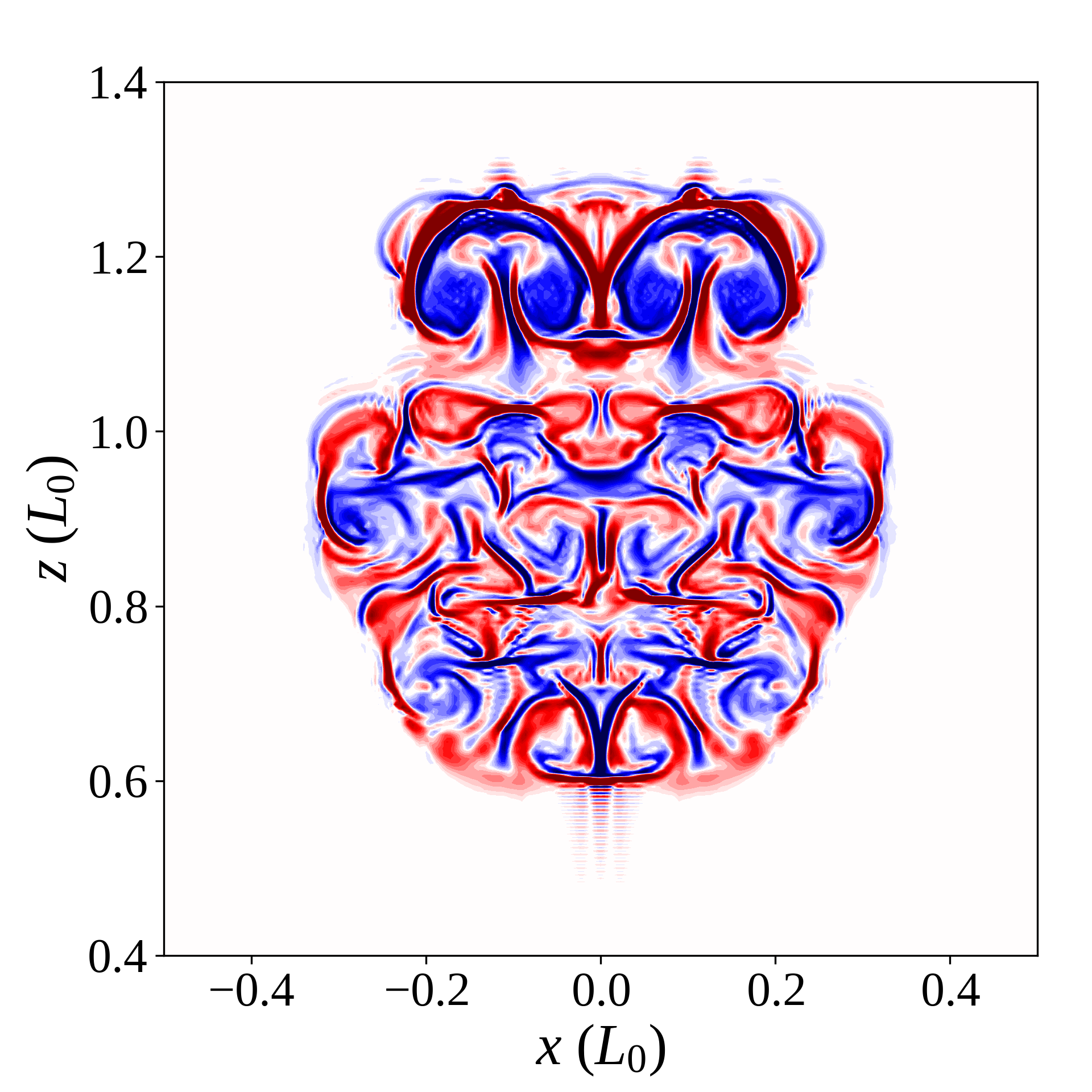}}%
	\subfloat[$t=380.0\,\tau_{\mathrm{A}}$.]{\label{fig:x7iJ_380}\includegraphics[keepaspectratio, width=0.305\linewidth, clip=true, trim=10 5 10 20]{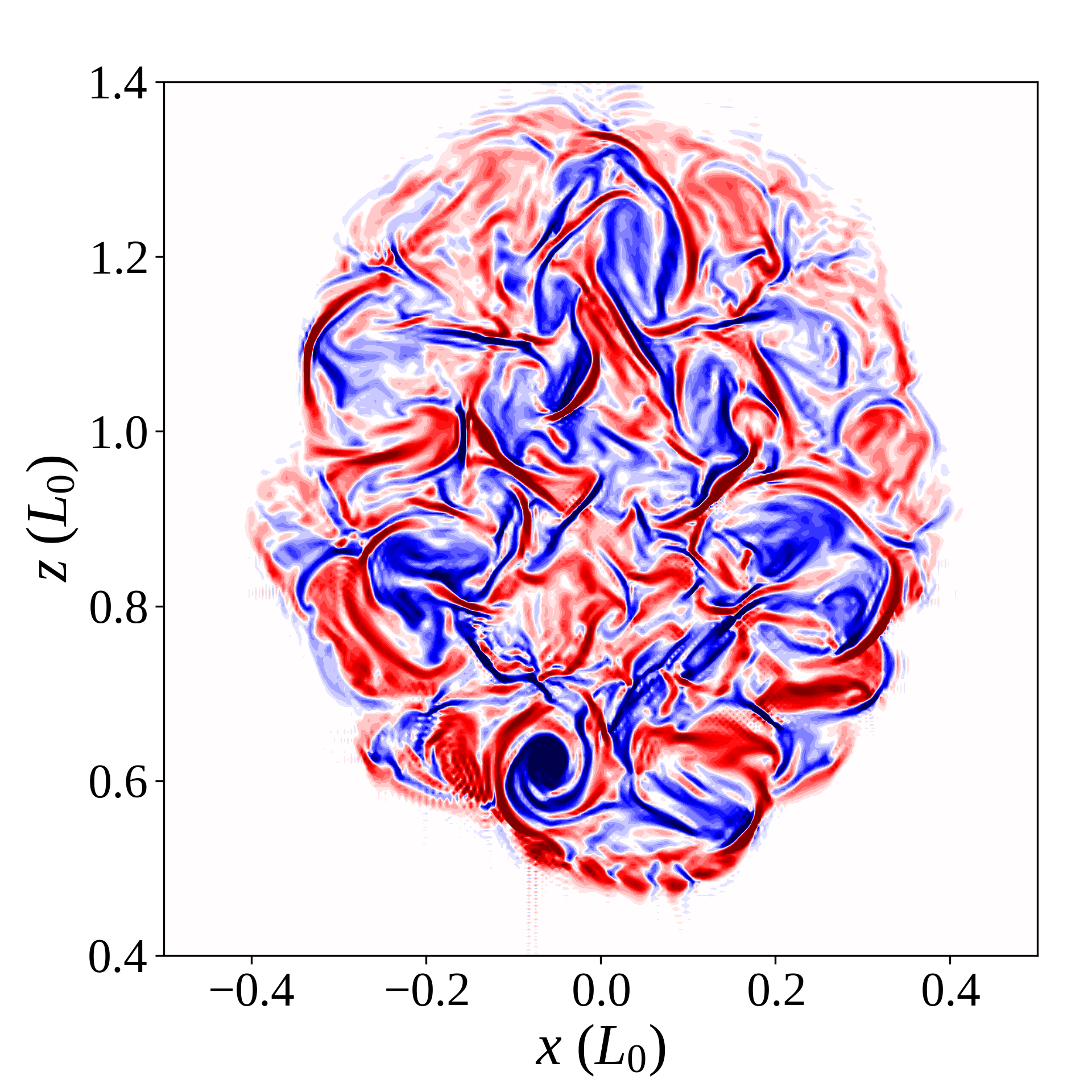}}%
	\includegraphics[keepaspectratio, width=0.085\linewidth, clip=true, trim=13 15 13 15]{JyContour_ColourBar.pdf}%
	\caption{Seven threads, identical drivers: Contours of toroidal current [$j_{y}$] above the polarity inversion line [$y=0$], at various times throughout the simulation (the key to contour levels is seen in colour bar).}
	\label{fig:x7iJ}
\end{figure*}

The 3D magnetic configuration and its evolution are illustrated in Figure~\ref{fig:x7iFL}; all seven threads are initially untwisted (Figure~\ref{fig:x7iFLa}), but they also travel higher into the domain than in the single-threaded case.
All threads are twisted at the same rate, but a large reconnection event (at the time shown in Figure~\ref{fig:x7iFLb}) causes the lowest row of threads to destabilise.
The times of disruption of each thread, and values of twist and decay index at those times, are recorded in Table~\ref{tab:threads}, for comparison with later experiments.
The destabilisation and fragmentation of the bottom row are followed very soon afterwards by the same in the middle row, with a short pause before the uppermost row of threads is disrupted (Figure~\ref{fig:x7iFLc}).
Another window on the evolution is provided by the energy and heating components of the system shown in Figure~\ref{fig:x7iEn}, coupled with contours of the current structures present at the apex of each of the threads (Figure~\ref{fig:x7iJ}).

As with the single-threaded case described in Section~\ref{sec:x1}, the system initially tends towards a potential state before the rotation commences.
All three components of energy rapidly increase following the onset of the driver, quickly forming seven threads.
The twist in field lines reaches the apex of the arcade before $t=140\,\tau_{\mathrm{A}}$: seven distinct current structures are visible in Figure~\ref{fig:x7iJ_140}.
The lower pair of threads become destabilised at/before $t=150\,\tau_{\mathrm{A}}$ (seen in Figure~\ref{fig:x7iJ_150}); this destabilisation occurs much earlier than in the single-threaded case.
Very shortly after this, threads in the middle row are all disrupted together (Figure~\ref{fig:x7iJ_157}), followed a short time later by the uppermost row of threads ($t=167.5\,\tau_{\mathrm{A}}$; Figure~\ref{fig:x7iJ_162}).
The time taken for all three rows of threads to destabilise is less than $20\,\tau_{\mathrm{A}}$ (compared with the complete experiment duration of $500\,\tau_{\mathrm{A}}$).
This concentrated and related series of disruptions is responsible for the largest bursts of Ohmic heating seen in Figure~\ref{fig:x7iEn} at $t=150\,\textrm{\textendash}\,170\,\tau_{\mathrm{A}}$.
However, these initial bursts significantly raise the internal energy of the system, to the extent that the change in internal energy becomes much larger than the change in magnetic energy.
One should note however that these trends are found upon subtracting the initial energy values; the plasma-$\beta$ is not materially affected.
This contrasts with Section~\ref{sec:x1}, where changes in magnetic and internal energy remained closely matched, but with a slightly larger change in the magnetic component throughout the single- threaded case.

Remnants of all seven threads now sporadically create thin current sheets, scattered throughout the volume and prone to reconnect; in particular, many of these later current sheets are associated with the lower and middle rows of threads.
At much later times (such as $t=380\,\tau_{\mathrm{A}}$ in Figure~\ref{fig:x7iJ_380}), the residue of current from all threads has expanded significantly, but is still pervaded by small current sheets, which occasionally, and irregularly, trigger anomalous resistivity.
Notably, the ongoing motions cause some threads apparently (and partially) to reform, as can be seen towards the bottom left of Figure~\ref{fig:x7iJ_380}.

Regarding the identification of a specific instability responsible for each disruption, Figure~\ref{fig:x7iJ} appears to suggest thread mergers (and not specific instabilities in individual threads) are responsible for many of the larger reconnection events and associated energy releases in this model.
We will further analyse the times, twist, and decay indices recorded in Table~\ref{tab:threads} in Section~\ref{sec:disc}, following a second experiment.

As in the single-threaded case, photospheric driving introduces a continuous Poynting flux from below and, in this case, marginally more of the injected energy ($\approx 58\,\%$) is ultimately converted to heat. Differences in number of threads, driving speeds, and footpoint areas between the cases further complicate a comparison of Poynting flux conversion between these cases.

\section{Seven Threads: Faster Central Thread} \label{sec:x7m}

\begin{figure*}[t]
	\includegraphics[keepaspectratio, width=\linewidth]{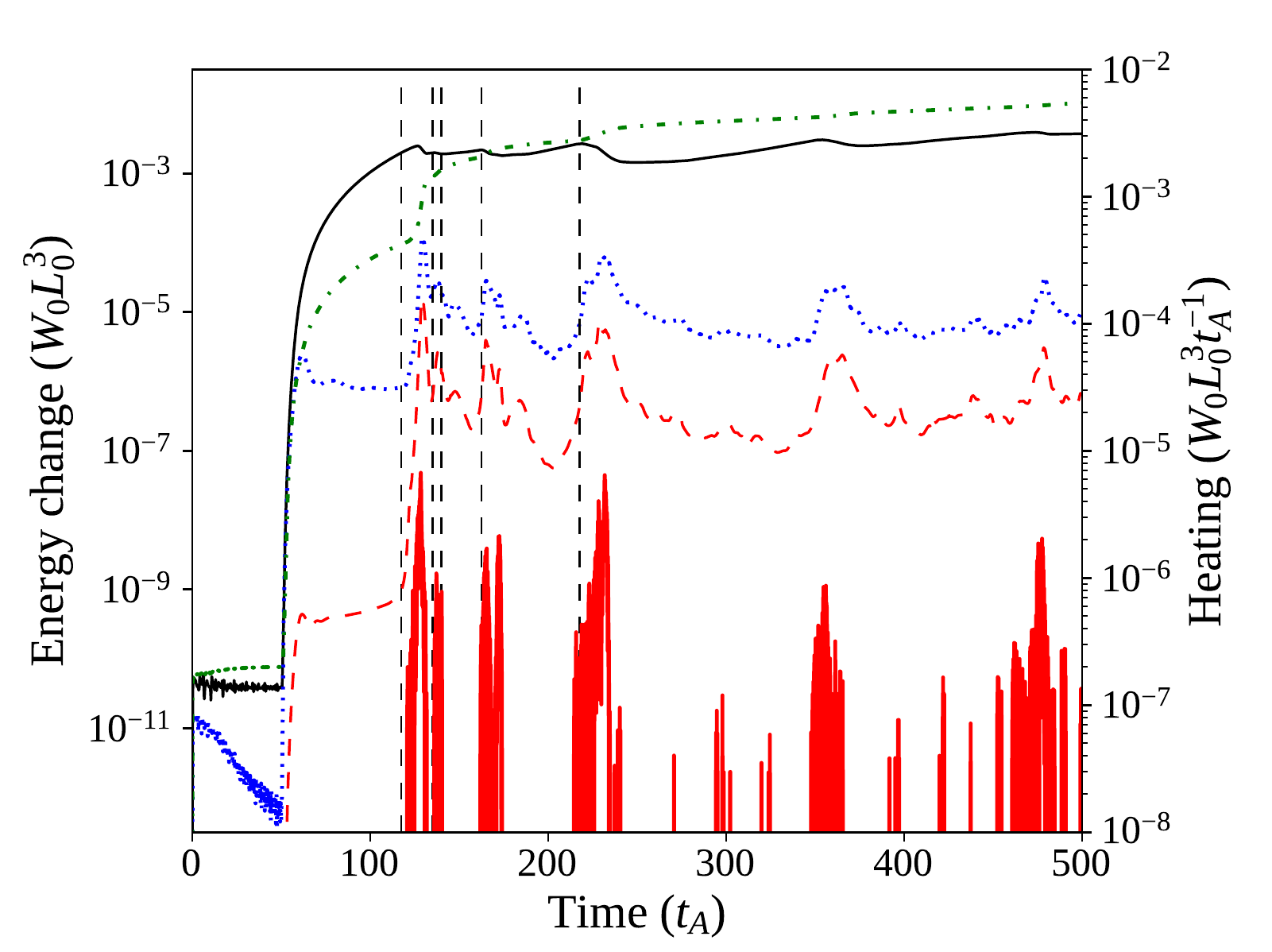}
	\caption{Seven threads, faster central thread, energies: Single-threaded case, energies: Magnetic (solid, black), internal (dash-dotted, green), and kinetic (dotted, blue) energy components with their initial values subtracted, shown together with Ohmic (thick, red) and viscous (dashed, red) instantaneous heating of the system.
	The six dashed, vertical lines indicate times of interest examined using contours of current in Figure~\ref{fig:x7mJ}.}
	\label{fig:x7mEn}
\end{figure*}

\begin{figure*}
	\centering
	\subfloat[$t=117.5\,\tau_{\mathrm{A}}$.]{\label{fig:x7mJ_117}\includegraphics[keepaspectratio, width=0.305\linewidth, clip=true, trim=10 5 10 20]{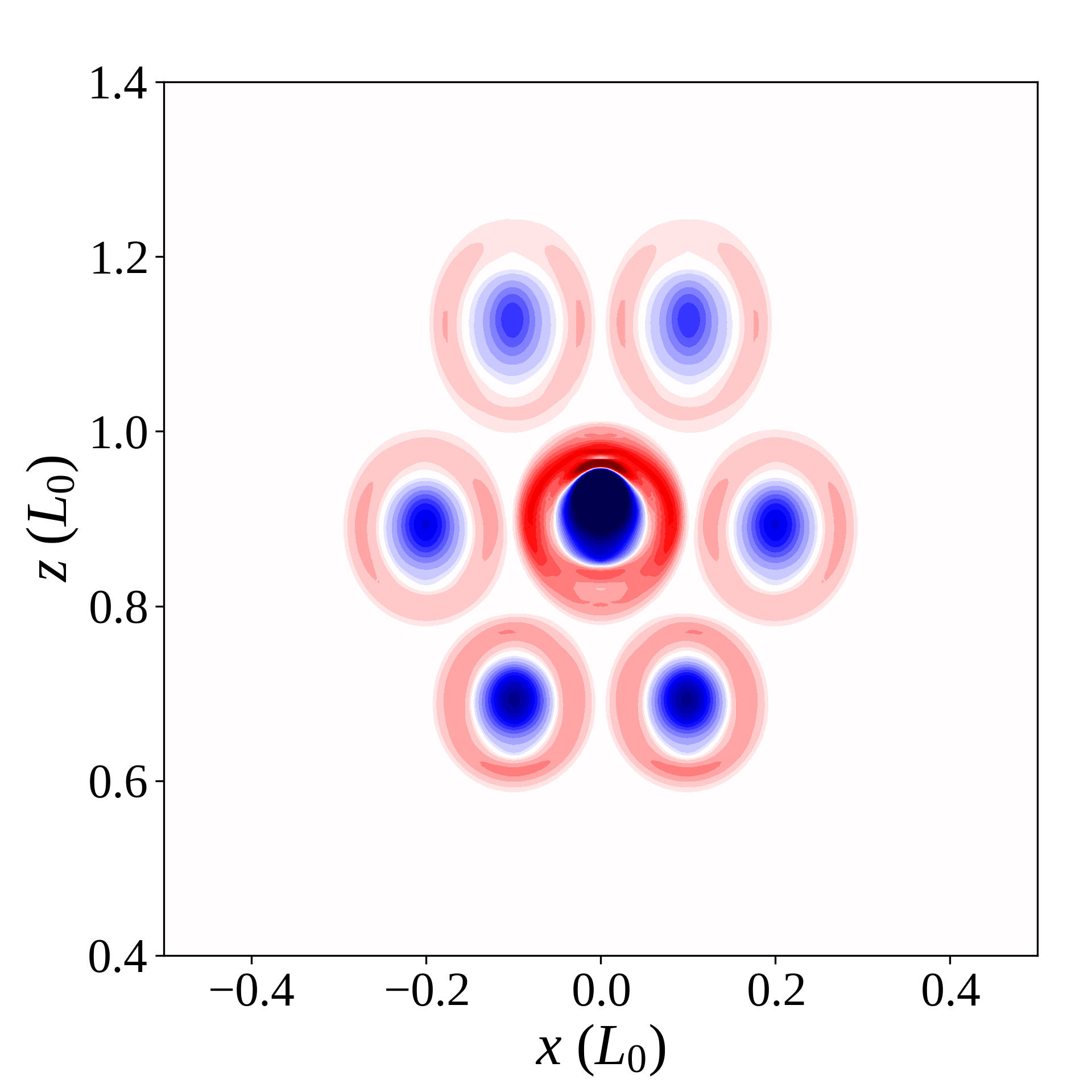}}%
	\subfloat[$t=135.0\,\tau_{\mathrm{A}}$.]{\label{fig:x7mJ_135}\includegraphics[keepaspectratio, width=0.305\linewidth, clip=true, trim=10 5 10 20]{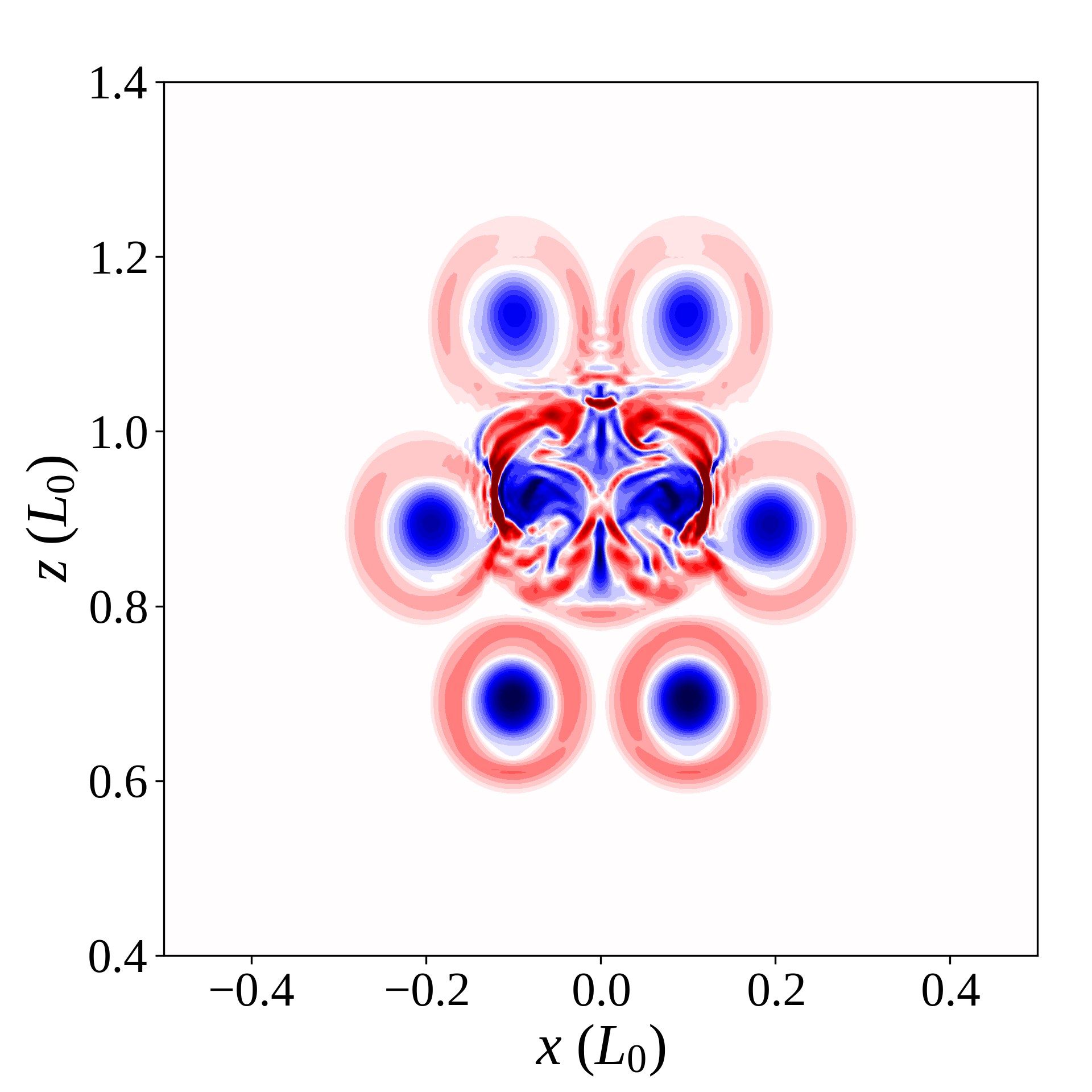}}%
	\subfloat[$t=140.0\,\tau_{\mathrm{A}}$.]{\label{fig:x7mJ_140}\includegraphics[keepaspectratio, width=0.305\linewidth, clip=true, trim=10 5 10 20]{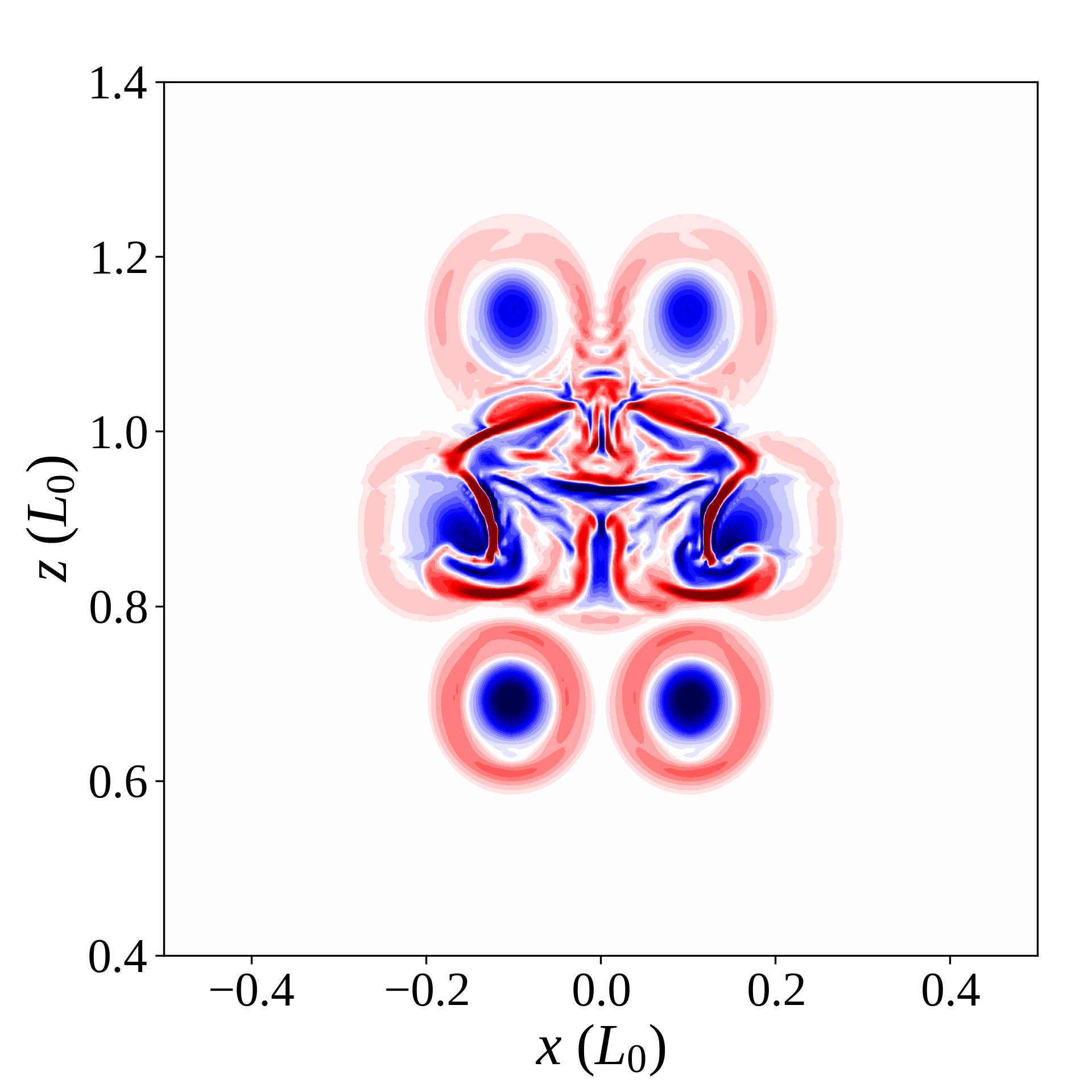}}%
	\includegraphics[keepaspectratio, width=0.085\linewidth, clip=true, trim=13 15 13 15]{JyContour_ColourBar.pdf}\\%
	\subfloat[$t=162.5\,\tau_{\mathrm{A}}$.]{\label{fig:x7mJ_162}\includegraphics[keepaspectratio, width=0.305\linewidth, clip=true, trim=10 5 10 20]{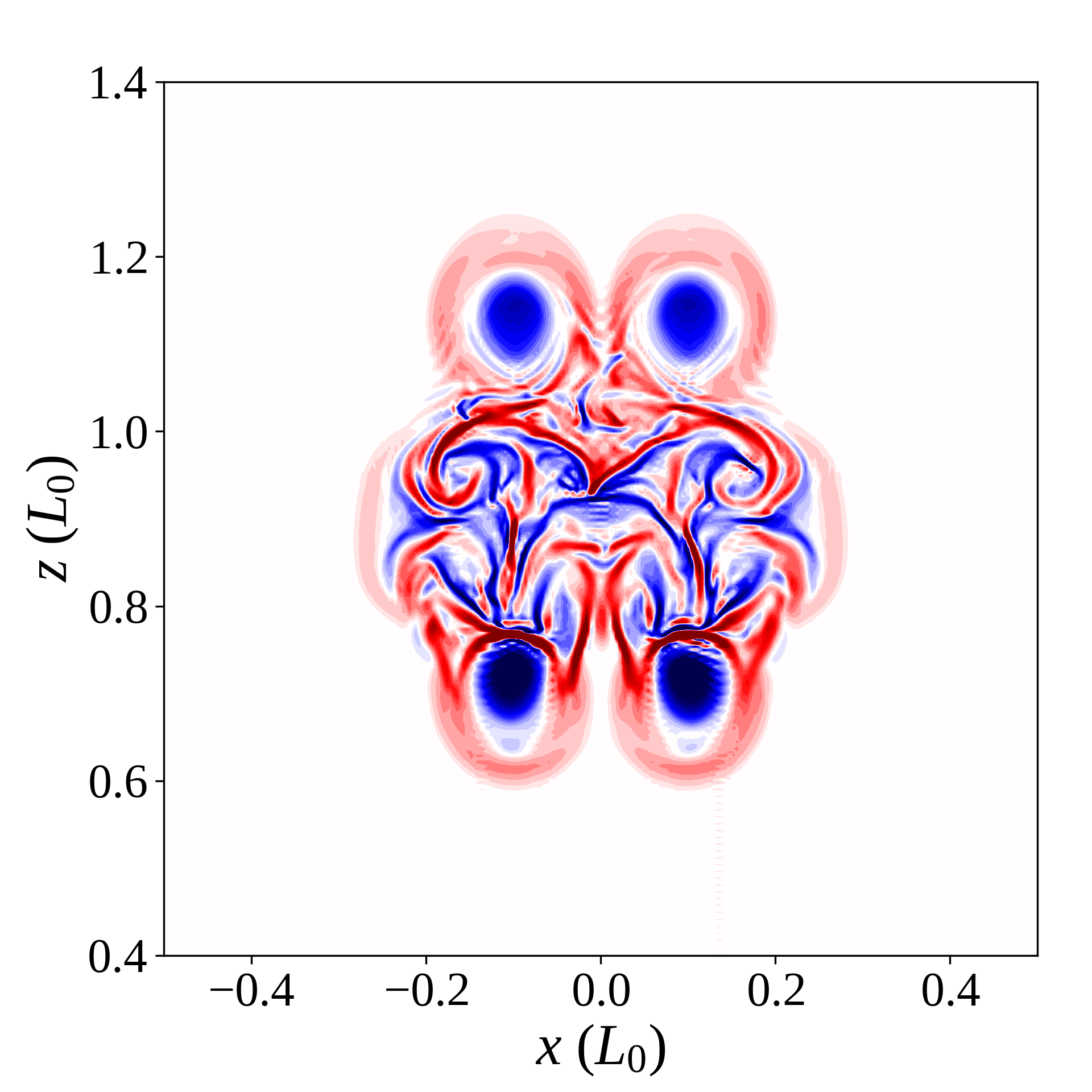}}%
	\subfloat[$t=217.5\,\tau_{\mathrm{A}}$.]{\label{fig:x7mJ_217}\includegraphics[keepaspectratio, width=0.305\linewidth, clip=true, trim=10 5 10 20]{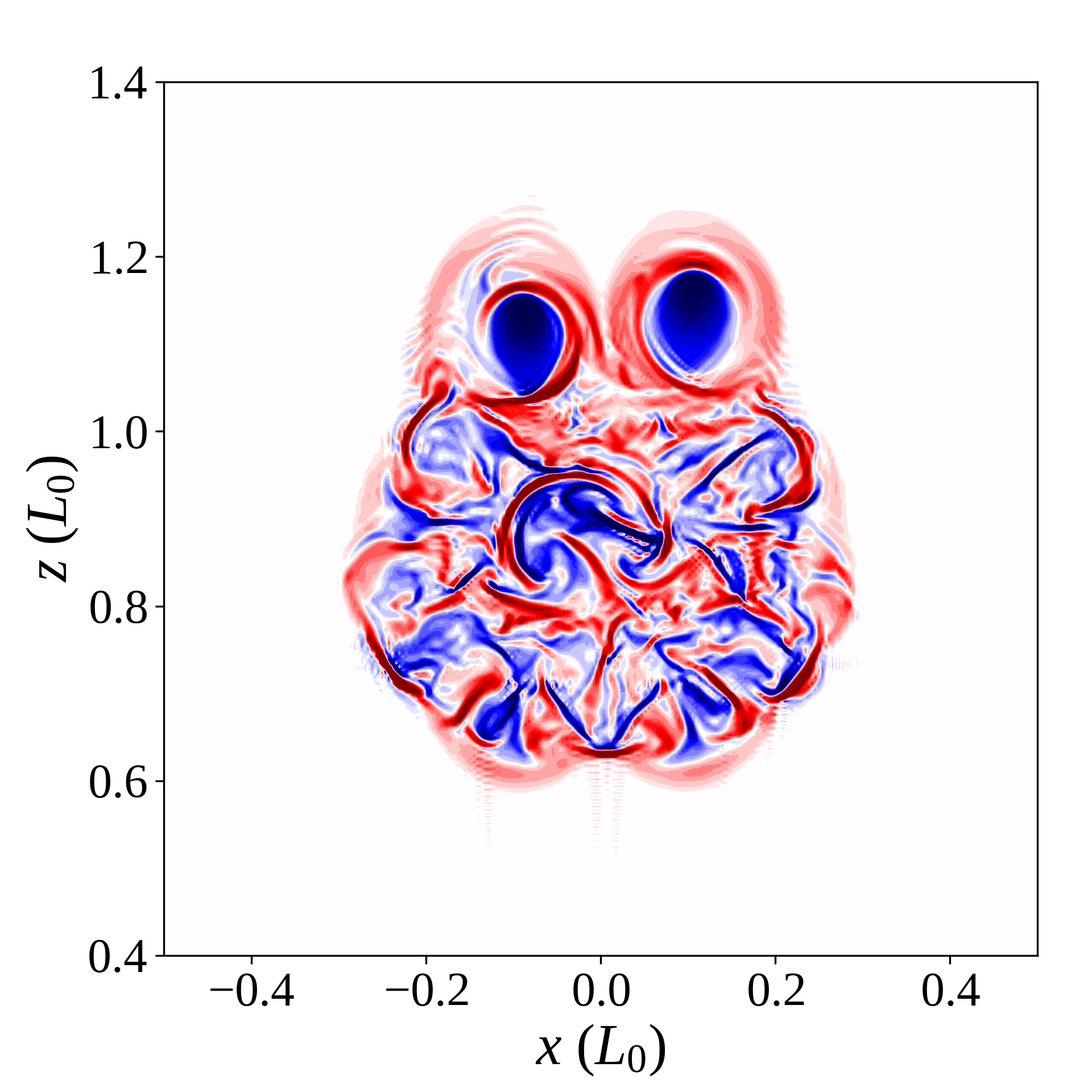}}%
	\subfloat[$t=500.0\,\tau_{\mathrm{A}}$.]{\label{fig:x7mJ_500}\includegraphics[keepaspectratio, width=0.305\linewidth, clip=true, trim=10 5 10 20]{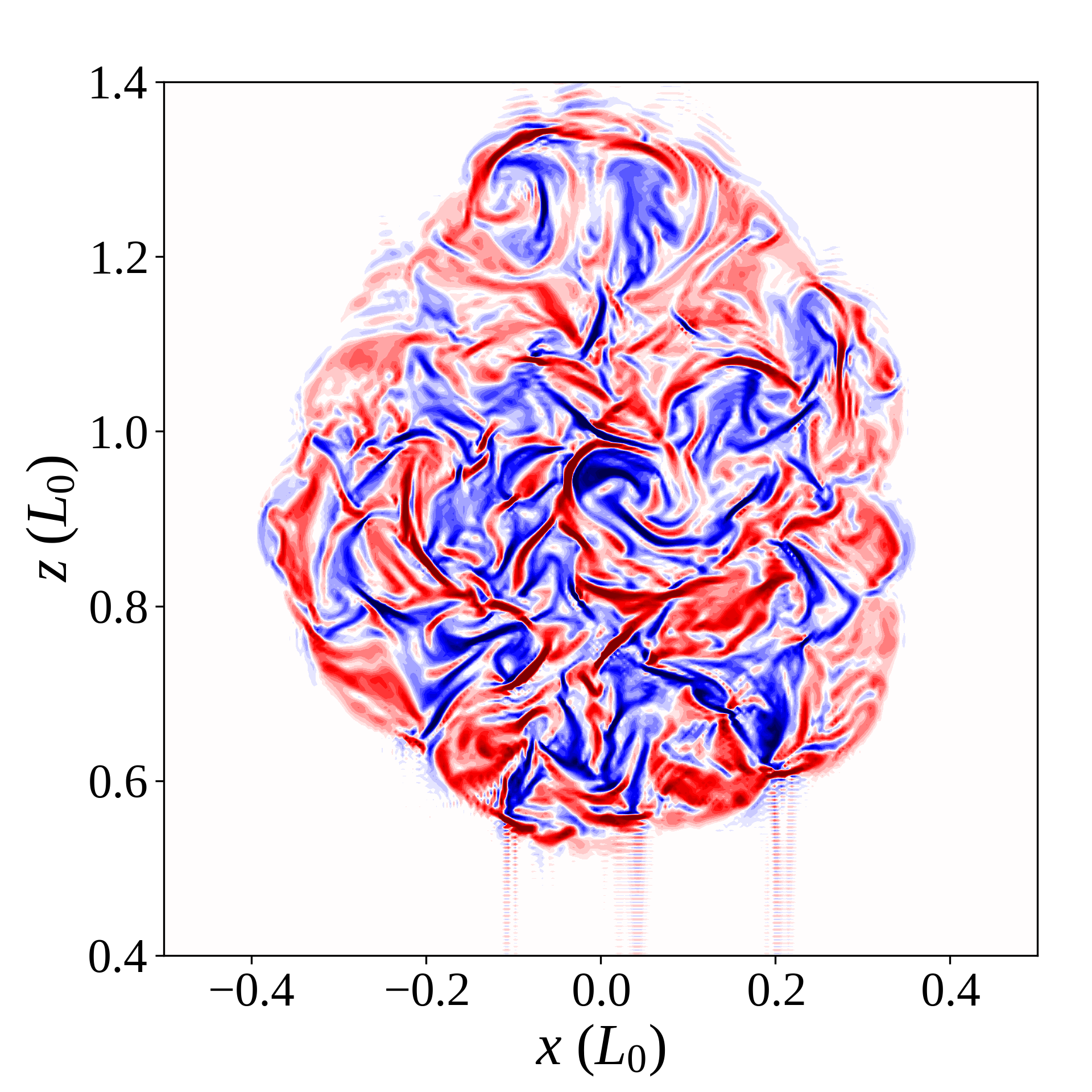}}%
	\includegraphics[keepaspectratio, width=0.085\linewidth, clip=true, trim=13 15 13 15]{JyContour_ColourBar.pdf}%
	\caption{Seven threads, central thread fastest: Contours of toroidal current [$j_y$,] above the polarity inversion line [$y=0$] at various times throughout the simulation (the key to contour levels is seen in colour bar).}
	\label{fig:x7mJ}
\end{figure*}

Our final investigation considers the effect of driving {speed} in individual threads; by enhancing the driving speed of the central thread, we {focus on} the role (if any) of variations {in speed} in the cascade process.
In the same configuration as in Section~\ref{sec:x7i}, we repeat the simulation with the central thread twisted three times faster ($v_{0}=0.03$) than the others ($v_{0}=0.01$).
Other parameters, including the relative positions of threads and resistivity conditions, remain unchanged.
The 3D evolution of the system is similar to that seen in Section~\ref{sec:x7i}; we will describe the evolution of this experiment using energetics and heating (in Figure~\ref{fig:x7mEn}) and through contours of current through the apex of the multi-threaded loop (in Figure~\ref{fig:x7mJ}).

The early evolution of the system, before $t\approx100\,\tau_{\mathrm{A}}$, largely matches that described in Section~\ref{sec:x7i} for threads driven at the same speed; a sufficiently potential state is realised before photospheric drivers create each thread.
Differences between multi-threaded experiments emerge soon after; current associated with the central thread builds up much faster than in surrounding threads (as shown, e.g., in Figure~\ref{fig:x7mJ_117}). 
The central thread is the first to destabilise (Figure~\ref{fig:x7mJ_135}).
This destabilisation is accompanied by a burst of Ohmic heating, enhanced internal energy, and decreased magnetic energy ($t\approx 130\,\tau_{\mathrm{A}}$, Figure~\ref{fig:x7mEn}).
Two more energy releases {soon} follow, as the remaining two threads in the middle row merge with the remnants of the central thread (Figure~\ref{fig:x7mJ_140}).
Once the middle row is entirely disrupted, there is a short pause ($\approx20\,\tau_{\mathrm{A}}$) in which little energy is dissipated or heating is seen.
At {$t\approx160\,\tau_{\mathrm{A}}$,} the remnants of the central row of threads begin to disrupt the lower row (Figure~\ref{fig:x7mJ_162}).
Two corresponding bursts of Ohmic heating are seen in Figure~\ref{fig:x7mEn} at approximately $t=165\,\tau_{\mathrm{A}}$ and $175\,\tau_{\mathrm{A}}$, respectively.
The two uppermost threads remain largely unaffected until $\approx220\,\tau_{\mathrm{A}}$, when they also begin to merge with the central row of threads (Figure~\ref{fig:x7mJ_217}), leading to yet another burst of Ohmic heating.
Between the disruption of this final pair of threads and the end of the simulation, further sporadic bursts of Ohmic heating continue to occur.
As before, these are driven by the continual photospheric driving, forming thin, fragmented current sheets throughout the expanding remnants of the threads.
Unlike in Section~\ref{sec:x7i}, these bursts appear more concentrated in time, with long intervals between larger bursts of Ohmic heating.
Although often insufficiently large (and hence unable to trigger anomalous resistive effects), several strong, thin fragmented current sheets are clearly visible in the final state of the experiment, seen in Figure~\ref{fig:x7mJ_500}.

Compared with the previous multi-threaded case (Section~\ref{sec:x7i}), the variation in driving speeds generates significant differences in the magnitude of Poynting flux injected.
Despite further differences in the avalanche process, a very similar overall proportion ($\approx 54\,\%$) of Poynting flux is ultimately converted into heat.
Once more, we record the approximate times of disruption of each thread, the associated mean twist, and the relevant values of the decay index, in Table~\ref{tab:threads}.

\begin{table}[t]	
	\begin{tabular}{crccccccc}
		\hline
		\multicolumn{3}{c}{Thread} &
		\multicolumn{3}{c}{Identical Drivers} &
		\multicolumn{3}{c}{Fast Central Thread} \\
		No. & Colour & Row & Time & $\langle\Phi\rangle$ & $n$ & Time & $\langle\Phi\rangle$ &$n$\\
		\hline
		1 & \cellcolor{Yellow}Yellow & Central & 155.0 & $3.41\,\pi$ & 1.308 & 120.0 & $4.12\,\pi$ & 1.272 \\
		2 & \cellcolor{Purple} \color{White}{Purple} & Central & 155.0 & $3.53\,\pi$ & 1.296 & 137.5 & $1.05\,\pi$ & 1.233 \\
		3 & \cellcolor{Navy} \color{White}{Navy} & Central & 155.0 & $3.13\,\pi$ & 1.295 & 137.5 & $1.17\,\pi$ & 1.233\\
		4 & \cellcolor{Cerulean} Cerulean & Upper & 162.5 & $3.49\,\pi$ & 1.515 & 222.5 & $3.41\,\pi$ & 1.466 \\
		5 & \cellcolor{SkyBlue} Sky Blue & Lower & 150.0 & $2.67\,\pi$ & 1.062 & 162.5 & $1.19\,\pi$ & 1.017 \\
		6 & \cellcolor{Orange} Orange & Upper & 162.5 & $3.49\,\pi$ & 1.515 & 217.5 & $2.54\,\pi$ & 1.466 \\
		7 & \cellcolor{Red} Red & Lower & 150.0 & $2.74\,\pi$ & 1.035 & 162.5 & $1.18\,\pi$ & 1.017 \\
		\hline
	\end{tabular}
	\caption{Seven threads: times of disruption, average twist $\langle\Phi\rangle$, and decay index $n$ for each thread in two experiments: one with all threads driven at same rate (Section~\ref{sec:x7i}) and one where the central thread is driven three times faster than surrounding threads (Section~\ref{sec:x7m}).
	Each thread is identified by a number and the assigned colour in 3D images (Figure~\ref{fig:x7iFL}) or field-line end-point locations (Figure~\ref{fig:x7FLend}).
	Times of disruption are discerned by eye from the contours of current in the $x$-$z$-plane at $y=0$.
	(As a consequence of the limited output cadence of simulation data, times of disruption are limited to specific discrete values. Other diagnostic information (for example the current in Figures~\ref{fig:x7iJ} and \ref{fig:x7mJ}) is only available at this lower cadence.
	Disruption times could alternatively be identified using (higher cadence) energy data, shown in Figures~\ref{fig:x7iEn} and \ref{fig:x7mEn}, distinguished by an exponential rise in the kinetic energy, associated with each instability.
	However, such rises are difficult to identify for subsequent disruptions in the overall energies of the system.)
	Twist is determined from field lines twisting helically around the axis of each thread, at the times of disruption.
	Decay indices quoted are evaluated at the intersection of the axis of each thread with the plane $y=0$ \citep[following the practice of][]{paper:Zuccarelloetal2015}.}
	\label{tab:threads}
\end{table}

\section{Analysis} \label{sec:disc}

Photospheric boundary motions, acting upon a curved magnetic arcade, have been shown to be capable of creating and supporting both single- and multi-threaded flux tubes.
The results of Sections~\ref{sec:x7i} and~\ref{sec:x7m} clearly illustrate that disruption in one thread may trigger an avalanche-like process, wherein all the threads in a multi-threaded tube are destabilised.
Our discussion will focus on a comparison of our multi-threaded cases and the nature of the initial instability.

One issue that warrants clarification is the relative timing of instability.
In the single-threaded experiment, the initial instability occurs much later ($t=330\,\tau_{\mathrm{A}}$) than in either multi-threaded case ($t=150\,\tau_{\mathrm{A}}$ in Section~\ref{sec:x7i}, $t=127.5\,\tau_{\mathrm{A}}$ in Section~\ref{sec:x7m}).
This is the result of different driving speeds: $v_{0}$ is much smaller in the single-threaded case ($v_{0}=0.008$) than in the multi-threaded cases (in Section~\ref{sec:x7i}, $v_{0}=0.02$; in Section~\ref{sec:x7m}, the central thread has $v_{0}=0.03$, the outer threads $v_{0}=0.01$).
As mentioned in the introduction, and by \citet{paper:Bownessetal2013}, successful MHD simulations require compressed timescales; while the slower speed here suffices for the first case, greater driving speeds were necessary to model the more complex, multi-threaded behaviour.
Faster speeds relative to the single-threaded case, by factors of $2.5$ and $3.75$ respectively, 
bring forward the time of the initial instability proportionately.
Hence, it occurs approximately three or four times {sooner (after the start of driving) in each multi-threaded case than in the single-threaded case
(respectively, after $100\,\tau_\mathrm{A}$ or $75\,\tau_\mathrm{A}$, compared with $280\,\tau_\mathrm{A}$)}.

\begin{figure*}
	\centering
	\subfloat[$t=0$]{\label{fig:x7iCNX_0}\includegraphics[keepaspectratio, width=0.295\linewidth, clip=true, trim=15 5 135 25]{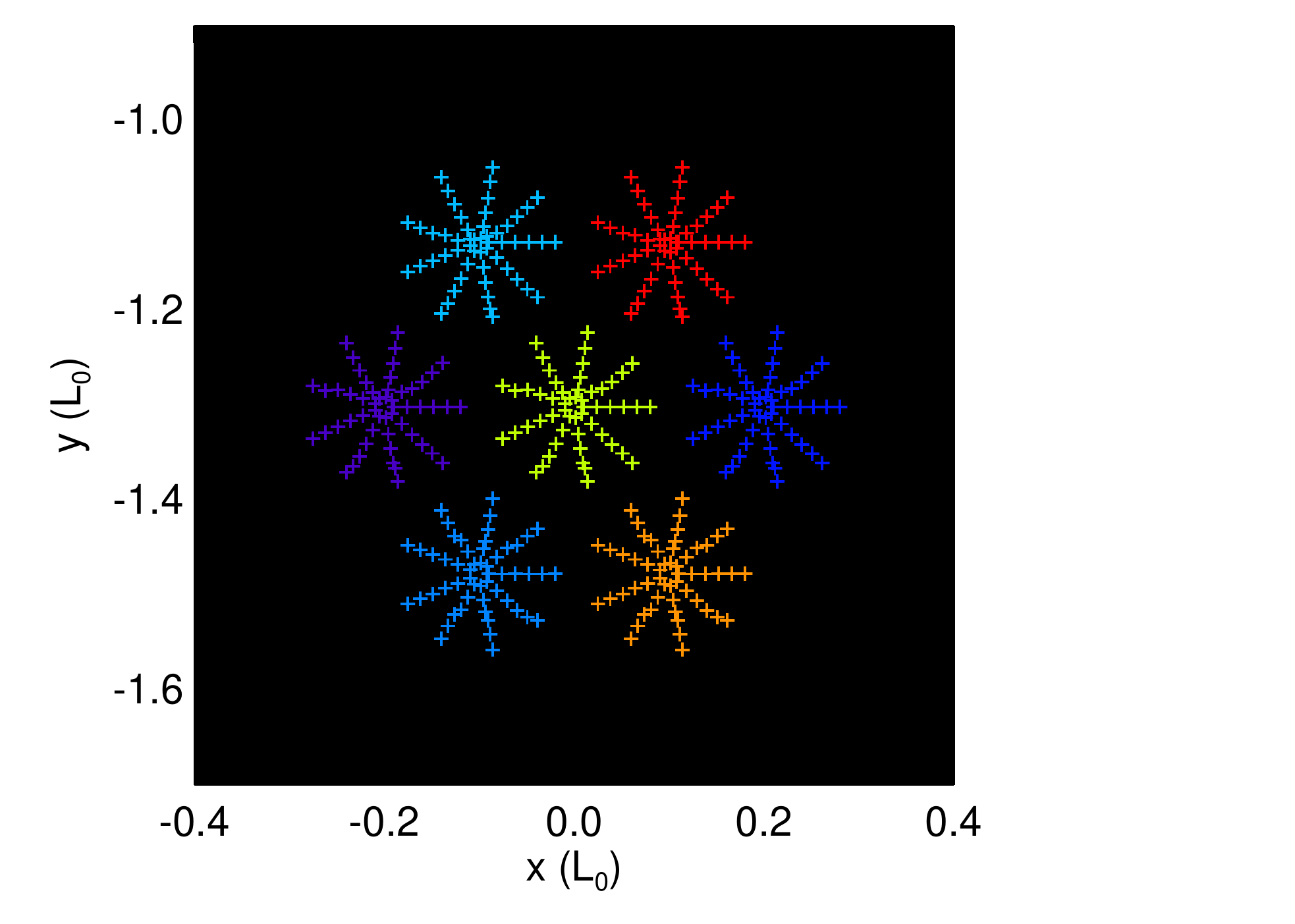}}%
	\subfloat[Identical speeds]{\label{fig:x7iCNX_500}\includegraphics[keepaspectratio, width=0.295\linewidth, clip=true, trim=15 5 135 25]{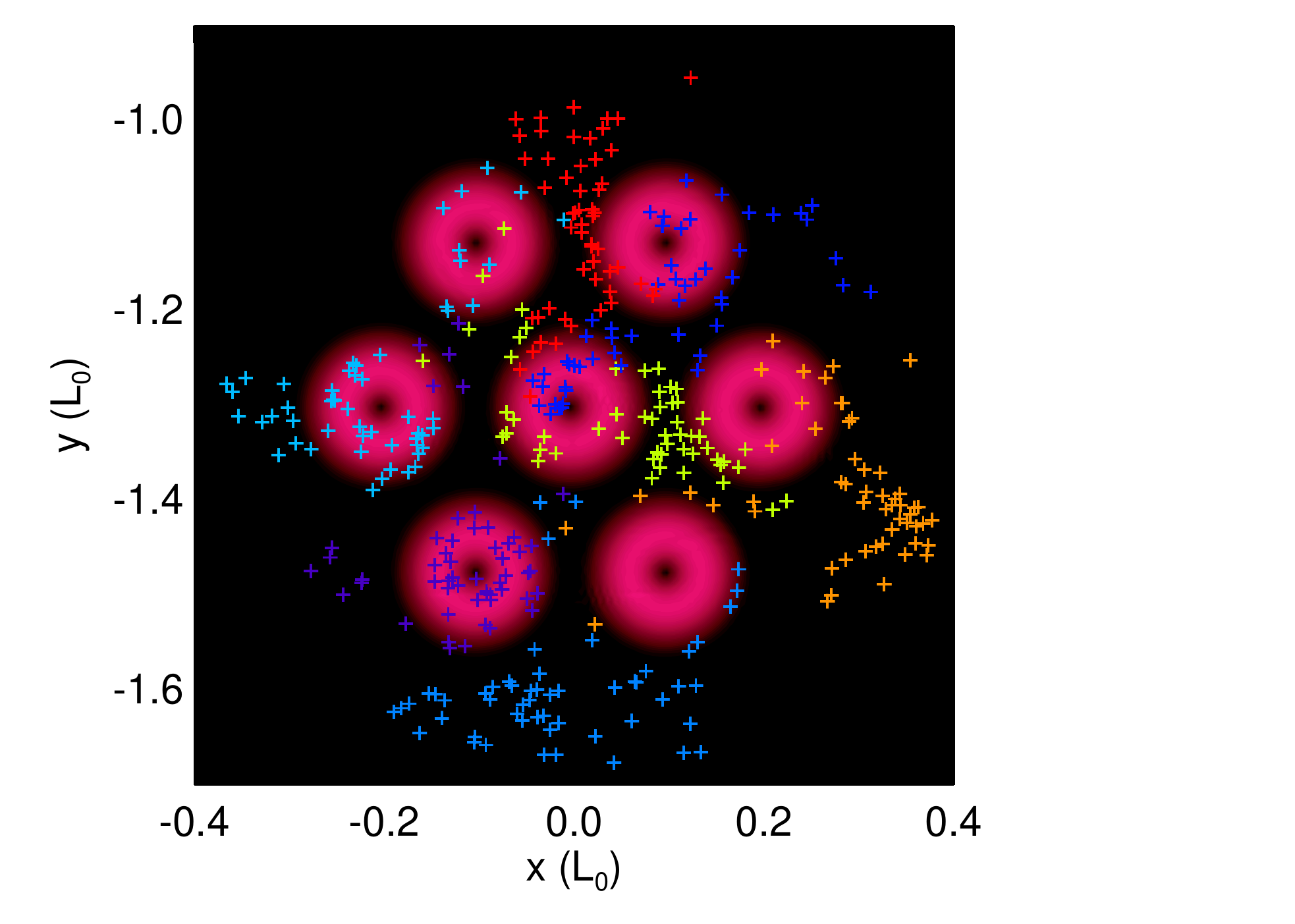}}%
	\subfloat[Faster central thread]{\label{fig:x7mCNX_500}\includegraphics[keepaspectratio, width=0.295\linewidth, clip=true, trim=15 5 135 25]{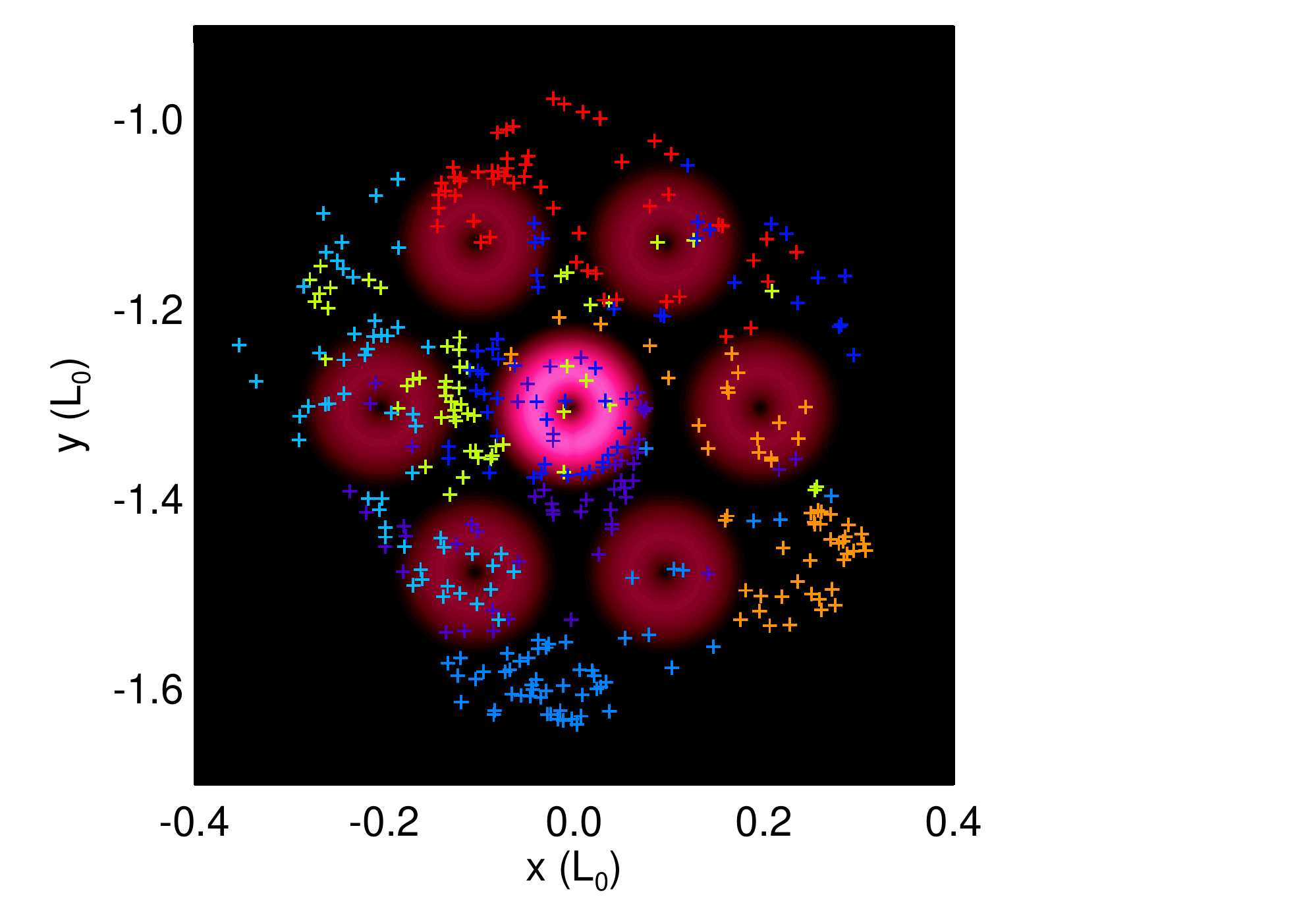}}%
	\includegraphics[keepaspectratio, width=0.115\linewidth, clip=true, trim=15 30 45 9]{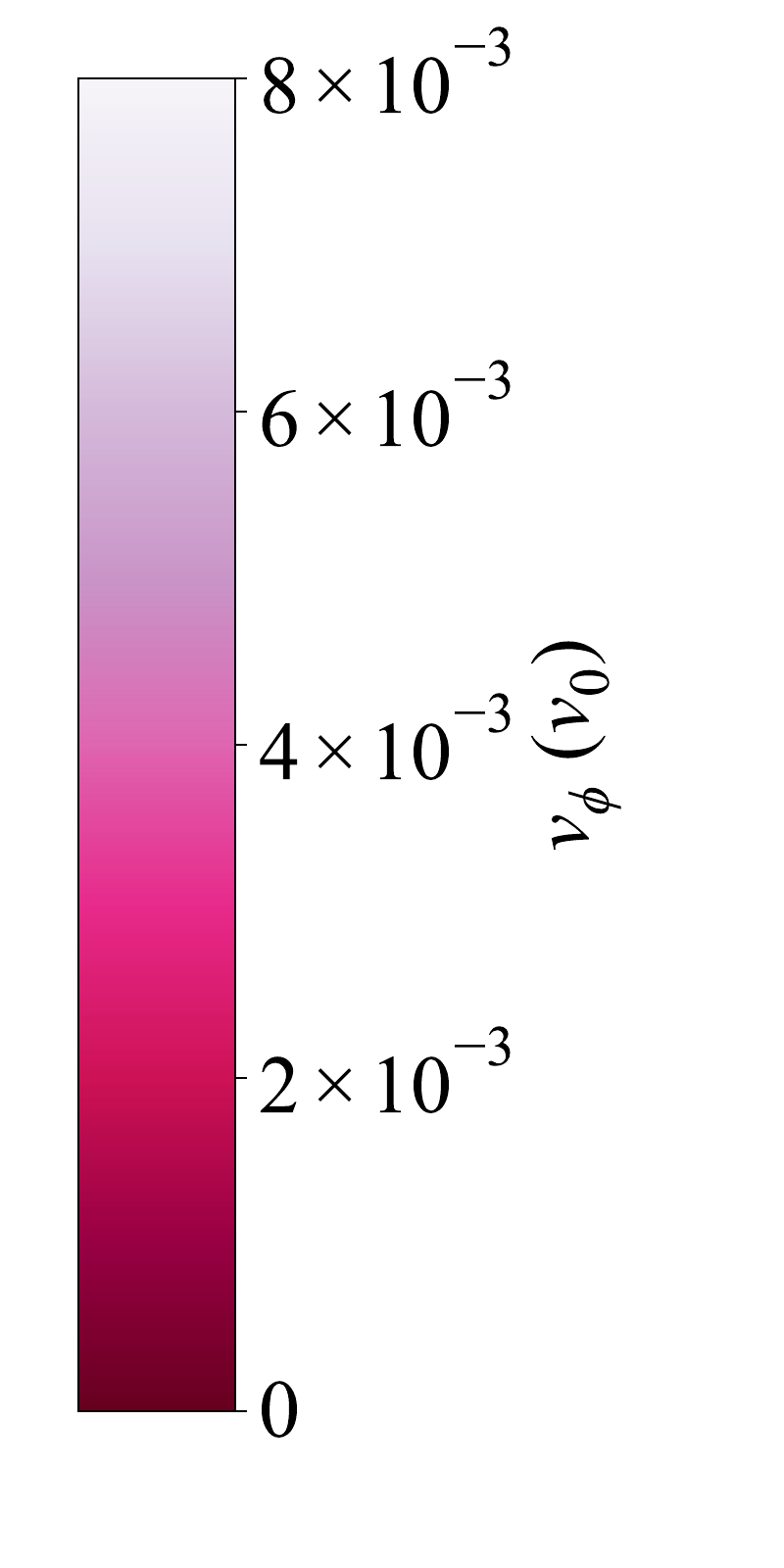}
	\caption{End-points of specific field lines within seven-threaded loop experiments.
Field lines are traced from footpoints at positive polarity to those at negative, coloured according to their initial position in a specific driving region or thread number (described in Table~\ref{tab:threads}) and overlaid onto the vortical driving profile at the base of the domain at the \protect\subref{fig:x7iCNX_0} beginning and \protect\subref{fig:x7iCNX_500} end of the uniform driven amplitude experiment, with \protect\subref{fig:x7mCNX_500} showing the final state of the faster central thread experiment.}
	\label{fig:x7FLend}
\end{figure*}

Comparing multi-threaded cases, both configurations ultimately undergo a very similar process and achieve end states that contain common traits.
One visual representation of this is the distribution of end positions of field lines starting within specific threads seen in Figure~\ref{fig:x7FLend}. 
The field lines associated with each thread become mixed with those from neighbouring threads by the end of the experiment in Figure~\ref{fig:x7FLend}.
The specific pattern in each experiment differs; this is to be expected when the threads destabilise and interact in a different order.
However, the remnants of all threads have expanded significantly, coalesced in the lanes created between photospheric driving regions, and ultimately fill much more of the volume of the simulation than when the driver was initiated.
Similar mixing is observed in straight cylindrical cases \citep[e.g.
Figure~7 of][]{paper:Reidetal2018}; with the arcade geometry, there is comparable, but perhaps slightly less widespread, entangling of flux.
However, seven threads in three rows may be insufficient to draw out any such influence: experiments containing more threads, or more rows of threads, may bring more evidence of geometrical effects. 

Our experiments recover a tendency of entire rows of threads (at the same height) to destabilise simultaneously.
Furthermore, the expansion of the field above the PIL also causes a significant gap between the uppermost and middle rows.
Separation attenuates the upward spread of instability and destabilization.
The arcade geometry of our model and variations in field strength with height lead to a preferential upward expansion near the apex of the loop.
These are clear effects imparted on the system evolution by our arcade geometry.

The preferential destabilisation of specific rows (with height above the PIL) may be influenced by other factors.
{Varying driving speed allows} for more direct comparisons with previous multi-threaded models \citep[e.g.][]{paper:Hoodetal2016}, wherein a single thread typically triggers further disruptions.
For a central thread driven three times faster than surrounding threads, the resulting avalanche proceeds differently from the case where all threads are driven at the same rate.
In Section~\ref{sec:x7m}, we recover a very similar picture to other multi-threaded models \citep[e.g. those of][]{paper:Reidetal2020a}, reporting that individual threads destabilise their neighbours in turn.
Despite some interaction between the middle and top rows, the top row destabilises largely as in the previous case.
The driving speed of threads on the top row is half that seen in Section~\ref{sec:x7i}, consistent with the merger of the top row occurring twice as long after the twisting starts.
Thus, it is possible to obtain at least a partial cascade of thread eruptions, but the avalanche process can be attenuated (or, indeed, suppressed entirely) by the field geometry.

The evidence for the nature of the initial instability (found in Table~\ref{tab:threads}) is mixed.
In the single-threaded case, the average twist {is} $\langle\Phi\rangle=1.68\,\pi$ when the thread destabilises.
This value is lower than in comparable investigations; our value is smaller than in the equivalent straight cylindrical case \citep[][]{paper:Reidetal2018}, and is nearly half as small as threshold values in models containing curvature and expansion \citep[$\Phi\approx3.26\,\pi$ in][]{paper:Gerrardetal2004} or line-tied loops \citep[$\Phi\approx 3.3\,\pi$, determined by][]{paper:HoodPriest1979b}.
Small values of twist at instability are less common, but not unheard of \citep[e.g.][ finding a mean twist $\approx\frac{\pi}{2}$]{paper:Rappazzoetal2019}.
\citet{paper:Barefordetal2016} assert that expansion and curvature of flux tubes significantly accelerate the onset of instability.
Our model contains more curvature and expansion than previous models, in order that thread footpoints are anchored in near-vertical field regions of the photosphere.
We also note that average twist is an ensemble measure.
Considering the peak twist, some field lines have twist around $7\,\pi$ at the onset of instability: this value far exceeds equivalent threshold values in the literature.
Turning to our multi-threaded configurations, in Section~\ref{sec:x7i} the lowest row of threads destabilise at $\langle\Phi\rangle=2.67\,\textrm{\textendash}\,2.74\,\pi$, while the central thread in Section~\ref{sec:x7m} achieves $\langle\Phi\rangle=4.12\,\pi$ at the point of instability.
These values are comparable with threshold values of \citet{paper:HoodPriest1979b,paper:Gerrardetal2004}, but more than more than a factor of two smaller than those of \citet{paper:Reidetal2018}.
The twist of subsequent threads at destabilisation depends on the experiment.
If all threads are driven at the same speed, threads which become unstable later are twisted more than the first.
If the central thread is driven faster, it becomes unstable first and then affects others; hence subsequent thread disruptions exhibit less twist.
The recorded values of twist imply that the kink instability is unlikely to be responsible for any later disruptions.

Turning to the torus instability, the value of $n$ at the apex in the single-threaded case ($n=1.572$) and the range over the cross-section are consistent with values commonly associated with the torus instability.
In the multi-threaded cases, values of $n$ are generally lower: the bottom-most threads in Section~\ref{sec:x7i} show values $n=1.04\,\textrm{\textendash}\,1.06$, lower than such typical values.
The faster central thread in Section~\ref{sec:x7m} achieves $n=1.272$ at disruption, again less indicative of instability.
Values of $n$ increase with height, making the uppermost threads more likely than the lower to be susceptible to the torus instability.
As a consequence of the superposition of Fourier modes in order to achieve near-vertical field at the footpoints, the loop here has a flatter, off-circular shape, evident in the red field lines of Figure~\ref{fig:Initial_x1FL}.
This is less favourable to such an instability, and the rise of the threads before instability is neither far nor sustained.
As such, and given the small net current, it is unlikely that the instability here is a conventional torus instability \citep{paper:Demoulinetal2010,paper:ReesCrockfordetal2020}.

Here, as elsewhere, it is seen that separating conjectured instabilities can be challenging.
For the spread of the avalanche, the most important facet is how an ideal instability in a single thread triggers the destabilisation of additional, stable threads.

One final issue with models of curved flux tubes relies upon our chosen velocity scales.
In Section~\ref{sec:x1}, our model assumes mean velocities $\approx8.12\times 10^{-4}\,v_{\mathrm{A}}$.
Given our normalisation (Table~\ref{tab:norm}), this corresponds physically to $561\,\mathrm{m}\,\mathrm{s}^{-1}$.
Typical velocity scales in simulations of photospheric velocity patterns often approach several tens or hundreds of metres per second \citep{review:GizonBirch2005,review:RieutordRincon2010}.
We approach realistic driving timescales in this model, in addition to our realistic magnetic structure.
Our MHD approach affords some flexibility in normalisation, allowing us to consider speeds even closer to those observed.

\section{Conclusions and Future Work} \label{sec:conc}

Our investigation means that we are now able to answer the question posed in the title of this work: multi-threaded curved flux tubes are indeed capable of supporting a magnetohydrodynamic avalanche.
Our magnetic arcade field has been shown to be capable of supporting flux tubes, formed through rotational photospheric driving.
As with straight cylindrical examples \citep[e.g.][]{paper:Reidetal2018}, continuous driving of individual threads leads to instability and the destabilisation of other stable threads, whereupon anomalous resistive effects release and redistribute the field lines of the entire tube.
The initial instabilities show characteristics common to both kink and torus instabilities.
Continual driving of (post-disruption) flux tube remnants leads to the formation of secondary current sheets, which cause aperiodic bursts of Ohmic heating, releasing more energy from the field.
Thus an arcade field may continue to yield bursty reconnection events when continuously driven from below.

Our findings diverge from similar models based on straight cylindrical geometry; our cases containing more than one thread demonstrate that the geometry of the magnetic field itself (here, a curved arcade structure) and the imposed driving speed of individual threads \emph{both} play a significant role in the formation and evolution of an MHD avalanche.
In the case where the speed is the same across all footpoints, threads preferentially destabilise through mergers with other threads of the same height and field strength, separate from the others; twisting all threads at the same rate in a single flux tube leads to a gradual destabilisation of threads, from strong to weak field strength.
However, driving a single thread unstable more quickly in a region of strong magnetic field can lead to a cascade of destabilisation of other threads, which more closely resembles previous examples between parallel planes; individual threads can destabilise neighbours, regardless of field strength.
The geometric configuration (and resulting variation in field strength) may also attenuate the cascade in some parts of the loop.

Several opportunities for further investigation readily present themselves.
Our curved model geometry provides an important first step towards a more realistic model of a magnetic arcade, and yet relies on a finite number of Fourier modes.
Each additional mode affects curvature and will hence potentially affect the mode of instability recovered: this merits further consideration.
Similarly, our geometry requires a novel critical parameter [$\zeta$] in order to trigger anomalous resistive effects upon specific current sheets: would the use of such a critical threshold alter/affect the findings seen in straight cylindrical cases \citep[e.g.][]{paper:Reidetal2020a}?

Looking ahead still further, this model still omits several relevant (yet potentially complex) physical factors that warrant consideration: these include the effect of gravity; a stratified atmosphere (in particular, stratifying density and temperature); and thermodynamic effects, such as thermal conduction and radiation.
Our model has generated many complex, highly time-dependent current sheets, distributed throughout the volume of the flux tube; further analysis of the distribution, lifetimes, and strengths of currents is warranted, together with examination of the impact of model parameters (including field strength and resistivity).
From these current sheets, heating is continuously being produced, the magnitude, composition, and distribution of which will be discussed in greater detail.
How do the electric fields associated with these current sheets compare with those observed or inferred solar observations, and might these electric fields be capable of accelerating particles?
Finally, our initial photospheric driving pattern is relatively simplistic: could a more complex photospheric driver create and/or drive flux tubes in this model? 

\begin{acknowledgements}
	The authors gratefully acknowledge the financial support of STFC through the Consolidated grant, ST/S000402/1, to the University of St Andrews.
	J. Threlfall is grateful for support from the Division of Games Technology and Mathematics at Abertay University.
	A.W. Hood acknowledges support from ERC Synergy grant \lq\lq The Whole Sun\rq\rq\ (810218).
	The authors are grateful to Vasilis Archontis for instructive advice and fruitful discussion concerning the manuscript.
	This work used the Python programming language \citep{vanRossum1991}, together with the NumPy \citep{Harris2020} and Matplotlib \citep{Hunter2007} packages.
	This work used
	the DIRAC 1, UKMHD Consortium machine at the University of St Andrews;
	the DiRAC Data Centric system at Durham University, operated by the Institute for Computational Cosmology;
	the DiRAC Data Analytic system at the University of Cambridge, operated by the University of Cambridge High Performance Computing Service; and
	the Cambridge Service for Data Driven Discovery (CSD3), which is operated by the University of Cambridge Research Computing.
	These systems are operated on behalf of the STFC DiRAC HPC Facility (\url{dirac.ac.uk}).
	The equipment was funded by
	National E-infrastructure capital grants (ST/K00042X/1 and ST/K001590/1),
	STFC capital grants (ST/K00087X/1, ST/H008861/1, ST/H00887X/1, ST/P002307/1, and ST/R002452/1),
	DiRAC Operations grants (ST/K003267/1 and ST/K00333X/1), and
	STFC operations grant ST/R00689X/1.
	DiRAC is part of the National E-Infrastructure.
\end{acknowledgements}

\section*{Disclosure of Potential Conflicts of Interests}
The authors declare that they have no conflicts of interest.

\bibliographystyle{spr-mp-sola}
\bibliography{SOLA-D-21-00066R2}

\begin{thebibliography}{47}
\ifx\bisbn     \undefined \def\bisbn  #1{ISBN #1}\fi
\ifx\binits    \undefined \def\binits#1{#1}\fi
\ifx\bauthor   \undefined \def\bauthor#1{#1}\fi
\ifx\batitle   \undefined \def\batitle#1{#1}\fi
\ifx\bjtitle   \undefined \def\bjtitle#1{\textit{#1}}\fi
\ifx\bvolume   \undefined \def\bvolume#1{\textbf{#1}}\fi
\ifx\byear     \undefined \def\byear#1{#1}\fi
\ifx\bissue    \undefined \def\bissue#1{#1}\fi
\ifx\bfpage    \undefined \def\bfpage#1{#1}\fi
\ifx\blpage    \undefined \def\blpage #1{#1}\fi
\ifx\burl      \undefined \def\burl#1{#1}\fi
\ifx\href      \undefined \def\href#1#2{#2}\fi
\ifx\betal     \undefined \def\betal{et al.}\fi
\ifx\bctitle   \undefined \def\bctitle#1{#1}\fi
\ifx\beditor   \undefined \def\beditor#1{#1}\fi
\ifx\bbtitle   \undefined \def\bbtitle#1{\textit{#1}}\fi
\ifx\bedition  \undefined \def\bedition#1{#1}\fi
\ifx\bseriesno \undefined \def\bseriesno#1{\textbf{#1}}\fi
\ifx\blocation \undefined \def\blocation#1{#1}\fi
\ifx\bsertitle \undefined \def\bsertitle#1{\textit{#1}}\fi
\ifx\bsnm      \undefined \def\bsnm#1{#1}\fi
\ifx\bsuffix   \undefined \def\bsuffix#1{#1}\fi
\ifx\bparticle \undefined \def\bparticle#1{#1}\fi
\ifx\barticle  \undefined \def\barticle#1{}\fi
\ifx\binstitute  \undefined \def\binstitute#1{#1}\fi
\ifx\bpublisher  \undefined \def\bpublisher#1{#1}\fi
\ifx\doiurl    \undefined \def\doiurl#1{\href{#1}{DOI}}\fi
\makeatletter
\def\safeHref#1#2#3{\in@{http}{#2}\ifin@\href{#2}{#3}\else\href{#1#2}{#3}\fi}
\makeatother
\ifx\adsurl    \undefined
  \def\adsurl#1{\safeHref{https://ui.adsabs.harvard.edu/abs/}{#1}{ADS}}\fi
\ifx\arxivurl  \undefined
  \def\arxivurl#1{\safeHref{http://arxiv.org/abs/}{#1}{arXiv}}\fi
\ifx\botherref \undefined \def\botherref#1{}\fi
\ifx\url       \undefined \def\url#1{#1}\fi
\ifx\bchapter  \undefined \def\bchapter#1{}\fi
\ifx\bbook     \undefined \def\bbook#1{}\fi
\ifx\bcomment  \undefined \def\bcomment#1{#1}\fi
\ifx\oauthor   \undefined \def\oauthor#1{#1}\fi
\ifx\citeauthoryear \undefined\def \citeauthoryear#1{#1}\fi
\def\endbibitem {}
\ifx\bconflocation  \undefined \def\bconflocation#1{#1} \fi

\bibitem[\protect\citeauthoryear{{Arber} et~al.}{2001}]{paper:LareXd2001}
\begin{barticle}
\bauthor{\bsnm{{Arber}}, \binits{T.D.}},
\bauthor{\bsnm{{Longbottom}}, \binits{A.W.}},
\bauthor{\bsnm{{Gerrard}}, \binits{C.L.}},
\bauthor{\bsnm{{Milne}}, \binits{A.M.}}:
\byear{2001},
\batitle{{A Staggered Grid, Lagrangian-Eulerian Remap Code for 3-D MHD
  Simulations}}.
\bjtitle{\jcp}
\bvolume{171},
\bfpage{151}.
\doiurl{https://doi.org/10.1006/jcph.2001.6780}.
\adsurl{2001JCoPh.171..151A}.
\end{barticle}
\endbibitem

\bibitem[\protect\citeauthoryear{{Aulanier}
  et~al.}{2010}]{paper:Aulanieretal2010}
\begin{barticle}
\bauthor{\bsnm{{Aulanier}}, \binits{G.}},
\bauthor{\bsnm{{T{\"o}r{\"o}k}}, \binits{T.}},
\bauthor{\bsnm{{D{\'e}moulin}}, \binits{P.}},
\bauthor{\bsnm{{DeLuca}}, \binits{E.E.}}:
\byear{2010},
\batitle{{Formation of Torus-Unstable Flux Ropes and Electric Currents in
  Erupting Sigmoids}}.
\bjtitle{\apj}
\bvolume{708},
\bfpage{314}.
\doiurl{https://doi.org/10.1088/0004-637X/708/1/314}.
\adsurl{2010ApJ...708..314A}.
\end{barticle}
\endbibitem

\bibitem[\protect\citeauthoryear{{Bareford}, {Browning}, and {van der
  Linden}}{2010}]{paper:Barefordetal2010}
\begin{barticle}
\bauthor{\bsnm{{Bareford}}, \binits{M.R.}},
\bauthor{\bsnm{{Browning}}, \binits{P.K.}},
\bauthor{\bsnm{{van der Linden}}, \binits{R.A.M.}}:
\byear{2010},
\batitle{{A nanoflare distribution generated by repeated relaxations triggered
  by kink instability}}.
\bjtitle{\aap}
\bvolume{521},
\bfpage{A70}.
\doiurl{https://doi.org/10.1051/0004-6361/201014067}.
\adsurl{2010A&A...521A..70B}.
\end{barticle}
\endbibitem

\bibitem[\protect\citeauthoryear{{Bareford}, {Hood}, and
  {Browning}}{2013}]{paper:Barefordetal2013}
\begin{barticle}
\bauthor{\bsnm{{Bareford}}, \binits{M.R.}},
\bauthor{\bsnm{{Hood}}, \binits{A.W.}},
\bauthor{\bsnm{{Browning}}, \binits{P.K.}}:
\byear{2013},
\batitle{{Coronal heating by the partial relaxation of twisted loops}}.
\bjtitle{\aap}
\bvolume{550},
\bfpage{A40}.
\doiurl{https://doi.org/10.1051/0004-6361/201219725}.
\adsurl{2013A\%26A...550A..40B}.
\end{barticle}
\endbibitem

\bibitem[\protect\citeauthoryear{{Bareford}
  et~al.}{2016}]{paper:Barefordetal2016}
\begin{barticle}
\bauthor{\bsnm{{Bareford}}, \binits{M.R.}},
\bauthor{\bsnm{{Gordovskyy}}, \binits{M.}},
\bauthor{\bsnm{{Browning}}, \binits{P.K.}},
\bauthor{\bsnm{{Hood}}, \binits{A.W.}}:
\byear{2016},
\batitle{{Energy Release in Driven Twisted Coronal Loops}}.
\bjtitle{\solphys}
\bvolume{291},
\bfpage{187}.
\doiurl{https://doi.org/10.1007/s11207-015-0824-7}.
\adsurl{2016SoPh..291..187B}.
\end{barticle}
\endbibitem

\bibitem[\protect\citeauthoryear{{Bateman}}{1978}]{book:Bateman1978}
\begin{bbook}
\bauthor{\bsnm{{Bateman}}, \binits{G.}}:
\byear{1978},
\bbtitle{{MHD instabilities}},
\bpublisher{{MIT}},
\blocation{{Cambridge, MA}}.
\adsurl{1978mit..book.....B}.
\end{bbook}
\endbibitem

\bibitem[\protect\citeauthoryear{{Botha}, {Arber}, and
  {Hood}}{2011}]{paper:Bothaetal2011}
\begin{barticle}
\bauthor{\bsnm{{Botha}}, \binits{G.J.J.}},
\bauthor{\bsnm{{Arber}}, \binits{T.D.}},
\bauthor{\bsnm{{Hood}}, \binits{A.W.}}:
\byear{2011},
\batitle{{Thermal conduction effects on the kink instability in coronal
  loops}}.
\bjtitle{\aap}
\bvolume{525},
\bfpage{A96}.
\doiurl{https://doi.org/10.1051/0004-6361/201015534}.
\adsurl{2011A&A...525A..96B}.
\end{barticle}
\endbibitem

\bibitem[\protect\citeauthoryear{{Botha}, {Arber}, and
  {Srivastava}}{2012}]{paper:Bothaetal2012}
\begin{barticle}
\bauthor{\bsnm{{Botha}}, \binits{G.J.J.}},
\bauthor{\bsnm{{Arber}}, \binits{T.D.}},
\bauthor{\bsnm{{Srivastava}}, \binits{A.K.}}:
\byear{2012},
\batitle{{Observational Signatures of the Coronal Kink Instability with Thermal
  Conduction}}.
\bjtitle{\apj}
\bvolume{745},
\bfpage{53}.
\doiurl{https://doi.org/10.1088/0004-637X/745/1/53}.
\adsurl{2012ApJ...745...53B}.
\end{barticle}
\endbibitem

\bibitem[\protect\citeauthoryear{{Bowness}, {Hood}, and
  {Parnell}}{2013}]{paper:Bownessetal2013}
\begin{barticle}
\bauthor{\bsnm{{Bowness}}, \binits{R.}},
\bauthor{\bsnm{{Hood}}, \binits{A.W.}},
\bauthor{\bsnm{{Parnell}}, \binits{C.E.}}:
\byear{2013},
\batitle{{Coronal heating and nanoflares: current sheet formation and
  heating}}.
\bjtitle{\aap}
\bvolume{560},
\bfpage{A89}.
\doiurl{https://doi.org/10.1051/0004-6361/201116652}.
\adsurl{2013A&A...560A..89B}.
\end{barticle}
\endbibitem

\bibitem[\protect\citeauthoryear{{Browning} and {Van der
  Linden}}{2003}]{paper:BrowningVanderLinden2003}
\begin{barticle}
\bauthor{\bsnm{{Browning}}, \binits{P.K.}},
\bauthor{\bsnm{{Van der Linden}}, \binits{R.A.M.}}:
\byear{2003},
\batitle{{Solar coronal heating by relaxation events}}.
\bjtitle{\aap}
\bvolume{400},
\bfpage{355}.
\doiurl{https://doi.org/10.1051/0004-6361:20021887}.
\adsurl{2003A\%26A...400..355B}.
\end{barticle}
\endbibitem

\bibitem[\protect\citeauthoryear{{Browning}
  et~al.}{2008}]{paper:Browningetal2008}
\begin{barticle}
\bauthor{\bsnm{{Browning}}, \binits{P.K.}},
\bauthor{\bsnm{{Gerrard}}, \binits{C.}},
\bauthor{\bsnm{{Hood}}, \binits{A.W.}},
\bauthor{\bsnm{{Kevis}}, \binits{R.}},
\bauthor{\bsnm{{van der Linden}}, \binits{R.A.M.}}:
\byear{2008},
\batitle{{Heating the corona by nanoflares: simulations of energy release
  triggered by a kink instability}}.
\bjtitle{\aap}
\bvolume{485},
\bfpage{837}.
\doiurl{https://doi.org/10.1051/0004-6361:20079192}.
\adsurl{2008A\%26A...485..837B}.
\end{barticle}
\endbibitem

\bibitem[\protect\citeauthoryear{{Cirtain}
  et~al.}{2013}]{paper:Cirtainetal2013}
\begin{barticle}
\bauthor{\bsnm{{Cirtain}}, \binits{J.W.}},
\bauthor{\bsnm{{Golub}}, \binits{L.}},
\bauthor{\bsnm{{Winebarger}}, \binits{A.R.}},
\bauthor{\bsnm{{de Pontieu}}, \binits{B.}},
\bauthor{\bsnm{{Kobayashi}}, \binits{K.}},
\bauthor{\bsnm{{Moore}}, \binits{R.L.}},
\bauthor{\bsnm{{Walsh}}, \binits{R.W.}},
\bauthor{\bsnm{{Korreck}}, \binits{K.E.}},
\bauthor{\bsnm{{Weber}}, \binits{M.}},
\bauthor{\bsnm{{McCauley}}, \binits{P.}},
\bauthor{\bsnm{{Title}}, \binits{A.}},
\bauthor{\bsnm{{Kuzin}}, \binits{S.}},
\bauthor{\bsnm{{Deforest}}, \binits{C.E.}}:
\byear{2013},
\batitle{Energy release in the solar corona from spatially resolved magnetic
  braids}.
\bjtitle{\nat}
\bvolume{493},
\bfpage{501}.
\doiurl{https://doi.org/10.1038/nature11772}.
\adsurl{2013Natur.493..501C}.
\end{barticle}
\endbibitem

\bibitem[\protect\citeauthoryear{{De Pontieu}
  et~al.}{2014}]{paper:DePontieuetal2014}
\begin{barticle}
\bauthor{\bsnm{{De Pontieu}}, \binits{B.}},
\bauthor{\bsnm{{Rouppe van der Voort}}, \binits{L.}},
\bauthor{\bsnm{{McIntosh}}, \binits{S.W.}},
\bauthor{\bsnm{{Pereira}}, \binits{T.M.D.}},
\bauthor{\bsnm{{Carlsson}}, \binits{M.}},
\bauthor{\bsnm{{Hansteen}}, \binits{V.}},
\bauthor{\bsnm{{Skogsrud}}, \binits{H.}},
\bauthor{\bsnm{{Lemen}}, \binits{J.}},
\bauthor{\bsnm{{Title}}, \binits{A.}},
\bauthor{\bsnm{{Boerner}}, \binits{P.}},
\bauthor{\bsnm{{Hurlburt}}, \binits{N.}},
\bauthor{\bsnm{{Tarbell}}, \binits{T.D.}},
\bauthor{\bsnm{{Wuelser}}, \binits{J.P.}},
\bauthor{\bsnm{{De Luca}}, \binits{E.E.}},
\bauthor{\bsnm{{Golub}}, \binits{L.}},
\bauthor{\bsnm{{McKillop}}, \binits{S.}},
\bauthor{\bsnm{{Reeves}}, \binits{K.}},
\bauthor{\bsnm{{Saar}}, \binits{S.}},
\bauthor{\bsnm{{Testa}}, \binits{P.}},
\bauthor{\bsnm{{Tian}}, \binits{H.}},
\bauthor{\bsnm{{Kankelborg}}, \binits{C.}},
\bauthor{\bsnm{{Jaeggli}}, \binits{S.}},
\bauthor{\bsnm{{Kleint}}, \binits{L.}},
\bauthor{\bsnm{{Martinez-Sykora}}, \binits{J.}}:
\byear{2014},
\batitle{On the prevalence of small-scale twist in the solar chromosphere and
  transition region}.
\bjtitle{Science}
\bvolume{346}.
\doiurl{https://doi.org/10.1126/science.1255732}.
\adsurl{2014Sci...346D.315D}.
\end{barticle}
\endbibitem

\bibitem[\protect\citeauthoryear{{Démoulin} and
  {Aulanier}}{2010}]{paper:Demoulinetal2010}
\begin{barticle}
\bauthor{\bsnm{{Démoulin}}, \binits{P.}},
\bauthor{\bsnm{{Aulanier}}, \binits{G.}}:
\byear{2010},
\batitle{{Criteria for Flux Rope Eruption: Non-equilibrium Versus Torus
  Instability}}.
\bjtitle{\apj}
\bvolume{718},
\bfpage{1388}.
\doiurl{https://doi.org/10.1088/0004-637X/718/2/1388}.
\adsurl{2010ApJ...718.1388D}.
\end{barticle}
\endbibitem

\bibitem[\protect\citeauthoryear{{Gerrard}, {Hood}, and
  {Brown}}{2004}]{paper:Gerrardetal2004}
\begin{barticle}
\bauthor{\bsnm{{Gerrard}}, \binits{C.L.}},
\bauthor{\bsnm{{Hood}}, \binits{A.W.}},
\bauthor{\bsnm{{Brown}}, \binits{D.S.}}:
\byear{2004},
\batitle{{Effect of photospheric twisting motions on a confined toroidal
  loop}}.
\bjtitle{\solphys}
\bvolume{222},
\bfpage{79}.
\doiurl{https://doi.org/10.1023/B:SOLA.0000036877.20077.42}.
\adsurl{2004SoPh..222...79G}.
\end{barticle}
\endbibitem

\bibitem[\protect\citeauthoryear{{Gizon} and
  {Birch}}{2005}]{review:GizonBirch2005}
\begin{barticle}
\bauthor{\bsnm{{Gizon}}, \binits{L.}},
\bauthor{\bsnm{{Birch}}, \binits{A.C.}}:
\byear{2005},
\batitle{Local helioseismology}.
\bjtitle{\lrsp}
\bvolume{2}.
\doiurl{https://doi.org/10.12942/lrsp-2005-6}.
\adsurl{2005LRSP....2....6G}.
\end{barticle}
\endbibitem

\bibitem[\protect\citeauthoryear{{Harris} et~al.}{2020}]{Harris2020}
\begin{barticle}
\bauthor{\bsnm{{Harris}}, \binits{C.R.}},
\bauthor{\bsnm{{Millman}}, \binits{K.J.}},
\bauthor{\bsnm{{van der Walt}}, \binits{S.J.}},
\bauthor{\bsnm{{Gommers}}, \binits{R.}},
\bauthor{\bsnm{{Virtanen}}, \binits{P.}},
\bauthor{\bsnm{{Cournapeau}}, \binits{D.}},
\bauthor{\bsnm{{Wieser}}, \binits{E.}},
\bauthor{\bsnm{{Taylor}}, \binits{J.}},
\bauthor{\bsnm{{Berg}}, \binits{S.}},
\bauthor{\bsnm{{Smith}}, \binits{N.J.}},
\bauthor{\bsnm{{Kern}}, \binits{R.}},
\bauthor{\bsnm{{Picus}}, \binits{M.}},
\bauthor{\bsnm{{Hoyer}}, \binits{S.}},
\bauthor{\bsnm{{van Kerkwijk}}, \binits{M.H.}},
\bauthor{\bsnm{{Brett}}, \binits{M.}},
\bauthor{\bsnm{{Haldane}}, \binits{A.}},
\bauthor{\bsnm{{del Río}}, \binits{J.F.}},
\bauthor{\bsnm{{Wiebe}}, \binits{M.}},
\bauthor{\bsnm{{Peterson}}, \binits{P.}},
\bauthor{\bsnm{{Gérard-Marchant}}, \binits{P.}},
\bauthor{\bsnm{{Sheppard}}, \binits{K.}},
\bauthor{\bsnm{{Reddy}}, \binits{T.}},
\bauthor{\bsnm{{Weckesser}}, \binits{W.}},
\bauthor{\bsnm{{Abbasi}}, \binits{H.}},
\bauthor{\bsnm{{Gohlke}}, \binits{C.}},
\bauthor{\bsnm{{Oliphant}}, \binits{T.E.}}:
\byear{2020},
\batitle{{Array programming with {NumPy}}}.
\bjtitle{\nat}
\bvolume{585},
\bfpage{357}.
\doiurl{https://doi.org/10.1038/s41586-020-2649-2}.
\end{barticle}
\endbibitem

\bibitem[\protect\citeauthoryear{{Haynes} and
  {Arber}}{2007}]{paper:HaynesArber2007}
\begin{barticle}
\bauthor{\bsnm{{Haynes}}, \binits{M.}},
\bauthor{\bsnm{{Arber}}, \binits{T.D.}}:
\byear{2007},
\batitle{{Observational properties of a kink unstable coronal loop}}.
\bjtitle{\aap}
\bvolume{467},
\bfpage{327}.
\doiurl{https://doi.org/10.1051/0004-6361:20066922}.
\adsurl{2007A&A...467..327H}.
\end{barticle}
\endbibitem

\bibitem[\protect\citeauthoryear{{Hood} and
  {Priest}}{1979}]{paper:HoodPriest1979b}
\begin{barticle}
\bauthor{\bsnm{{Hood}}, \binits{A.W.}},
\bauthor{\bsnm{{Priest}}, \binits{E.R.}}:
\byear{1979},
\batitle{{Kink instability of solar coronal loops as the cause of solar
  flares}}.
\bjtitle{\solphys}
\bvolume{64},
\bfpage{303}.
\doiurl{https://doi.org/10.1007/BF00151441}.
\adsurl{1979SoPh...64..303H}.
\end{barticle}
\endbibitem

\bibitem[\protect\citeauthoryear{{Hood}, {Browning}, and {van der
  Linden}}{2009}]{paper:Hoodetal2009}
\begin{barticle}
\bauthor{\bsnm{{Hood}}, \binits{A.W.}},
\bauthor{\bsnm{{Browning}}, \binits{P.K.}},
\bauthor{\bsnm{{van der Linden}}, \binits{R.A.M.}}:
\byear{2009},
\batitle{{Coronal heating by magnetic reconnection in loops with zero net
  current}}.
\bjtitle{\aap}
\bvolume{506},
\bfpage{913}.
\doiurl{https://doi.org/10.1051/0004-6361/200912285}.
\adsurl{2009A\%26A...506..913H}.
\end{barticle}
\endbibitem

\bibitem[\protect\citeauthoryear{{Hood} et~al.}{2016}]{paper:Hoodetal2016}
\begin{barticle}
\bauthor{\bsnm{{Hood}}, \binits{A.W.}},
\bauthor{\bsnm{{Cargill}}, \binits{P.J.}},
\bauthor{\bsnm{{Browning}}, \binits{P.K.}},
\bauthor{\bsnm{{Tam}}, \binits{K.V.}}:
\byear{2016},
\batitle{{An MHD Avalanche in a Multi-threaded Coronal Loop.}}
\bjtitle{\apj}
\bvolume{817},
\bfpage{5}.
\doiurl{https://doi.org/10.3847/0004-637X/817/1/5}.
\adsurl{2016ApJ...817....5H}.
\end{barticle}
\endbibitem

\bibitem[\protect\citeauthoryear{{Howson}, {De Moortel}, and
  {Fyfe}}{2020}]{paper:Howsonetal2020}
\begin{barticle}
\bauthor{\bsnm{{Howson}}, \binits{T.A.}},
\bauthor{\bsnm{{De Moortel}}, \binits{I.}},
\bauthor{\bsnm{{Fyfe}}, \binits{L.E.}}:
\byear{2020},
\batitle{{The effects of driving time scales on heating in a coronal arcade}}.
\bjtitle{\aap}
\bvolume{643},
\bfpage{A85}.
\doiurl{https://doi.org/10.1051/0004-6361/202038869}.
\adsurl{2020A&A...643A..85H}.
\end{barticle}
\endbibitem

\bibitem[\protect\citeauthoryear{{Hunter}}{2007}]{Hunter2007}
\begin{barticle}
\bauthor{\bsnm{{Hunter}}, \binits{J.D.}}:
\byear{2007},
\batitle{{Matplotlib: A 2D Graphics Environment}}.
\bjtitle{Comput. Science Eng.}
\bvolume{9},
\bfpage{90}.
\burl{\url{ieeexplore.ieee.org/document/4160265}}.
\end{barticle}
\endbibitem

\bibitem[\protect\citeauthoryear{{Hussain}, {Browning}, and
  {Hood}}{2017}]{paper:Hussainetal2017}
\begin{barticle}
\bauthor{\bsnm{{Hussain}}, \binits{A.S.}},
\bauthor{\bsnm{{Browning}}, \binits{P.K.}},
\bauthor{\bsnm{{Hood}}, \binits{A.W.}}:
\byear{2017},
\batitle{{A relaxation model of coronal heating in multiple interacting flux
  ropes}}.
\bjtitle{\aap}
\bvolume{600},
\bfpage{A5}.
\doiurl{https://doi.org/10.1051/0004-6361/201629589}.
\adsurl{2017A\%26A...600A...5H}.
\end{barticle}
\endbibitem

\bibitem[\protect\citeauthoryear{{Kliem} and
  {T{\"o}r{\"o}k}}{2006}]{paper:KliemTorok2006}
\begin{barticle}
\bauthor{\bsnm{{Kliem}}, \binits{B.}},
\bauthor{\bsnm{{T{\"o}r{\"o}k}}, \binits{T.}}:
\byear{2006},
\batitle{{Torus Instability}}.
\bjtitle{\prl}
\bvolume{96},
\bfpage{255002}.
\doiurl{https://doi.org/10.1103/PhysRevLett.96.255002}.
\adsurl{2006PhRvL..96y5002K}.
\end{barticle}
\endbibitem

\bibitem[\protect\citeauthoryear{{O'Hara} and {De
  Moortel}}{2016}]{paper:OHaraDeMoortel2016}
\begin{barticle}
\bauthor{\bsnm{{O'Hara}}, \binits{J.P.}},
\bauthor{\bsnm{{De Moortel}}, \binits{I.}}:
\byear{2016},
\batitle{{Impact of flux distribution on elementary heating events}}.
\bjtitle{\aap}
\bvolume{594},
\bfpage{A67}.
\doiurl{https://doi.org/10.1051/0004-6361/201628913}.
\adsurl{2016A&A...594A..67O}.
\end{barticle}
\endbibitem

\bibitem[\protect\citeauthoryear{{Parker}}{1972}]{paper:Parker1972}
\begin{barticle}
\bauthor{\bsnm{{Parker}}, \binits{E.N.}}:
\byear{1972},
\batitle{{Topological Dissipation and the Small-Scale Fields in Turbulent
  Gases}}.
\bjtitle{\apj}
\bvolume{174},
\bfpage{499}.
\doiurl{https://doi.org/10.1086/151512}.
\adsurl{1972ApJ...174..499P}.
\end{barticle}
\endbibitem

\bibitem[\protect\citeauthoryear{{Parker}}{1988}]{paper:Parker1988}
\begin{barticle}
\bauthor{\bsnm{{Parker}}, \binits{E.N.}}:
\byear{1988},
\batitle{{Nanoflares and the solar X-ray corona}}.
\bjtitle{\apj}
\bvolume{330},
\bfpage{474}.
\doiurl{https://doi.org/10.1086/166485}.
\adsurl{1988ApJ...330..474P}.
\end{barticle}
\endbibitem

\bibitem[\protect\citeauthoryear{{Parnell} and {De
  Moortel}}{2012}]{review:ParnellDeMoortel2012}
\begin{barticle}
\bauthor{\bsnm{{Parnell}}, \binits{C.E.}},
\bauthor{\bsnm{{De Moortel}}, \binits{I.}}:
\byear{2012},
\batitle{{A contemporary view of coronal heating}}.
\bjtitle{\ptrsa}
\bvolume{370},
\bfpage{3217}.
\doiurl{https://doi.org/10.1098/rsta.2012.0113}.
\adsurl{2012RSPTA.370.3217P}.
\end{barticle}
\endbibitem

\bibitem[\protect\citeauthoryear{{Pinto}, {Vilmer}, and
  {Brun}}{2015}]{paper:Pintoetal2015}
\begin{barticle}
\bauthor{\bsnm{{Pinto}}, \binits{R.F.}},
\bauthor{\bsnm{{Vilmer}}, \binits{N.}},
\bauthor{\bsnm{{Brun}}, \binits{A.S.}}:
\byear{2015},
\batitle{{Soft X-ray emission in kink-unstable coronal loops}}.
\bjtitle{\aap}
\bvolume{576},
\bfpage{A37}.
\doiurl{https://doi.org/10.1051/0004-6361/201323358}.
\adsurl{2015A&A...576A..37P}.
\end{barticle}
\endbibitem

\bibitem[\protect\citeauthoryear{{Priest} and
  {Forbes}}{2000}]{book:PriestForbes}
\begin{bbook}
\bauthor{\bsnm{{Priest}}, \binits{E.R.}},
\bauthor{\bsnm{{Forbes}}, \binits{T.}}:
\byear{2000},
\bbtitle{{Magnetic Reconnection: MHD Theory and Applications}},
\bpublisher{Cambridge Univ. Press},
\blocation{New York}.
\bisbn{0521481791}.
\end{bbook}
\endbibitem

\bibitem[\protect\citeauthoryear{Rappazzo
  et~al.}{2019}]{paper:Rappazzoetal2019}
\begin{barticle}
\bauthor{\bsnm{Rappazzo}, \binits{A.F.}},
\bauthor{\bsnm{Velli}, \binits{M.}},
\bauthor{\bsnm{Dahlburg}, \binits{R.B.}},
\bauthor{\bsnm{Einaudi}, \binits{G.}}:
\byear{2019},
\batitle{Magnetic field line twisting by photospheric vortices: Energy storage
  and release}.
\bjtitle{\apj}
\bvolume{883},
\bfpage{148}.
\doiurl{https://doi.org/10.3847/1538-4357/ab3c69}.
\end{barticle}
\endbibitem

\bibitem[\protect\citeauthoryear{{Reale} et~al.}{2016}]{paper:Realeetal2016}
\begin{barticle}
\bauthor{\bsnm{{Reale}}, \binits{F.}},
\bauthor{\bsnm{{Orlando}}, \binits{S.}},
\bauthor{\bsnm{{Guarrasi}}, \binits{M.}},
\bauthor{\bsnm{{Mignone}}, \binits{A.}},
\bauthor{\bsnm{{Peres}}, \binits{G.}},
\bauthor{\bsnm{{Hood}}, \binits{A.W.}},
\bauthor{\bsnm{{Priest}}, \binits{E.R.}}:
\byear{2016},
\batitle{{3D MHD modeling of twisted coronal loops}}.
\bjtitle{\apj}
\bvolume{830},
\bfpage{21}.
\doiurl{https://doi.org/10.3847/0004-637X/830/1/21}.
\adsurl{2016ApJ...830...21R}.
\end{barticle}
\endbibitem

\bibitem[\protect\citeauthoryear{{Rees-Crockford}
  et~al.}{2020}]{paper:ReesCrockfordetal2020}
\begin{barticle}
\bauthor{\bsnm{{Rees-Crockford}}, \binits{T.}},
\bauthor{\bsnm{S.}, \binits{B.D.}},
\bauthor{\bsnm{{Scullion}}, \binits{E.}},
\bauthor{\bsnm{{Park}}, \binits{S.-H.}}:
\byear{2020},
\batitle{{2D and 3D Analysis of a Torus-unstable Quiet-Sun Prominence
  Eruption}}.
\bjtitle{\apj}
\bvolume{897},
\bfpage{35}.
\doiurl{https://doi.org/10.3847/1538-4357/ab92a0}.
\adsurl{2020ApJ...897...35R}.
\end{barticle}
\endbibitem

\bibitem[\protect\citeauthoryear{{Reid}}{2020}]{thesis:Reid}
\begin{botherref}
\oauthor{\bsnm{{Reid}}, \binits{J.}}:
2020,
A model for solar flares and coronal heating based on magnetohydrodynamic
  avalanches.
PhD thesis,
University of St Andrews.
\doiurl{https://doi.org/10.17630/sta/30}.
\adsurl{2020PhDT........13R}.
\url{research-repository.st-andrews.ac.uk/handle/10023/21489}.
\end{botherref}
\endbibitem

\bibitem[\protect\citeauthoryear{{Reid} et~al.}{2018}]{paper:Reidetal2018}
\begin{barticle}
\bauthor{\bsnm{{Reid}}, \binits{J.}},
\bauthor{\bsnm{{Hood}}, \binits{A.W.}},
\bauthor{\bsnm{{Parnell}}, \binits{C.E.}},
\bauthor{\bsnm{{Browning}}, \binits{P.K.}},
\bauthor{\bsnm{{Cargill}}, \binits{P.J.}}:
\byear{2018},
\batitle{{Coronal energy release by MHD avalanches: continuous driving}}.
\bjtitle{\aap}
\bvolume{615},
\bfpage{A84}.
\doiurl{https://doi.org/10.1051/0004-6361/201732399}.
\adsurl{2018A&A...615A..84R}.
\end{barticle}
\endbibitem

\bibitem[\protect\citeauthoryear{{Reid} et~al.}{2020}]{paper:Reidetal2020a}
\begin{barticle}
\bauthor{\bsnm{{Reid}}, \binits{J.}},
\bauthor{\bsnm{{Cargill}}, \binits{P.J.}},
\bauthor{\bsnm{{Hood}}, \binits{A.W.}},
\bauthor{\bsnm{{Parnell}}, \binits{C.E.}},
\bauthor{\bsnm{{Arber}}, \binits{T.D.}}:
\byear{2020},
\batitle{{Coronal energy release by MHD avalanches: Heating mechanisms}}.
\bjtitle{\aap}
\bvolume{633},
\bfpage{A158}.
\doiurl{https://doi.org/10.1051/0004-6361/201937051}.
\adsurl{2020A&A...633A.158R}.
\end{barticle}
\endbibitem

\bibitem[\protect\citeauthoryear{{Rieutord} and
  {Rincon}}{2010}]{review:RieutordRincon2010}
\begin{barticle}
\bauthor{\bsnm{{Rieutord}}, \binits{M.}},
\bauthor{\bsnm{{Rincon}}, \binits{F.}}:
\byear{2010},
\batitle{The sun's supergranulation}.
\bjtitle{\lrsp}
\bvolume{7}.
\doiurl{https://doi.org/10.12942/lrsp-2010-2}.
\adsurl{2010LRSP....7....2R}.
\end{barticle}
\endbibitem

\bibitem[\protect\citeauthoryear{{Snow} et~al.}{2017}]{paper:Snowetal2017}
\begin{barticle}
\bauthor{\bsnm{{Snow}}, \binits{B.}},
\bauthor{\bsnm{{Botha}}, \binits{G.J.J.}},
\bauthor{\bsnm{{R{\'e}gnier}}, \binits{S.}},
\bauthor{\bsnm{{Morton}}, \binits{R.J.}},
\bauthor{\bsnm{{Verwichte}}, \binits{E.}},
\bauthor{\bsnm{{Young}}, \binits{P.R.}}:
\byear{2017},
\batitle{{Observational Signatures of a Kink-unstable Coronal Flux Rope Using
  Hinode/EIS}}.
\bjtitle{\apj}
\bvolume{842},
\bfpage{16}.
\doiurl{https://doi.org/10.3847/1538-4357/aa6d0e}.
\adsurl{2017ApJ...842...16S}.
\end{barticle}
\endbibitem

\bibitem[\protect\citeauthoryear{{Syntelis}, {Archontis}, and
  {Tsinganos}}{2017}]{paper:Syntelisetal2017}
\begin{barticle}
\bauthor{\bsnm{{Syntelis}}, \binits{P.}},
\bauthor{\bsnm{{Archontis}}, \binits{V.}},
\bauthor{\bsnm{{Tsinganos}}, \binits{K.}}:
\byear{2017},
\batitle{{Recurrent CME-like Eruptions in Emerging Flux Regions. I. On the
  Mechanism of Eruptions}}.
\bjtitle{\apj}
\bvolume{850},
\bfpage{95}.
\doiurl{https://doi.org/10.3847/1538-4357/aa9612}.
\adsurl{2017ApJ...850...95S}.
\end{barticle}
\endbibitem

\bibitem[\protect\citeauthoryear{{Tam} et~al.}{2015}]{paper:Tametal2015}
\begin{barticle}
\bauthor{\bsnm{{Tam}}, \binits{K.V.}},
\bauthor{\bsnm{{Hood}}, \binits{A.W.}},
\bauthor{\bsnm{{Browning}}, \binits{P.K.}},
\bauthor{\bsnm{{Cargill}}, \binits{P.J.}}:
\byear{2015},
\batitle{{Coronal heating in multiple magnetic threads}}.
\bjtitle{\aap}
\bvolume{580},
\bfpage{A122}.
\doiurl{https://doi.org/10.1051/0004-6361/201525995}.
\adsurl{2015A\%26A...580A.122T}.
\end{barticle}
\endbibitem

\bibitem[\protect\citeauthoryear{{Threlfall}, {Hood}, and
  {Browning}}{2018}]{paper:Threlfalletal2018a}
\begin{barticle}
\bauthor{\bsnm{{Threlfall}}, \binits{J.}},
\bauthor{\bsnm{{Hood}}, \binits{A.W.}},
\bauthor{\bsnm{{Browning}}, \binits{P.K.}}:
\byear{2018},
\batitle{{Flare particle acceleration in the interaction of twisted coronal
  flux ropes}}.
\bjtitle{\aap}
\bvolume{611},
\bfpage{A40}.
\doiurl{https://doi.org/10.1051/0004-6361/201731915}.
\adsurl{2018A&A...611A..40T}.
\end{barticle}
\endbibitem

\bibitem[\protect\citeauthoryear{{Threlfall}, {Hood}, and
  {Priest}}{2018}]{paper:Threlfalletal2018b}
\begin{barticle}
\bauthor{\bsnm{{Threlfall}}, \binits{J.}},
\bauthor{\bsnm{{Hood}}, \binits{A.W.}},
\bauthor{\bsnm{{Priest}}, \binits{E.R.}}:
\byear{2018},
\batitle{{Flux Rope Formation Due to Shearing and Zipper Reconnection}}.
\bjtitle{\solphys}
\bvolume{293},
\bfpage{98}.
\doiurl{https://doi.org/10.1007/s11207-018-1318-1}.
\adsurl{2018SoPh..293...98T}.
\end{barticle}
\endbibitem

\bibitem[\protect\citeauthoryear{{Threlfall}, {Wright}, and
  {Hood}}{2020}]{paper:Threlfalletal2020}
\begin{barticle}
\bauthor{\bsnm{{Threlfall}}, \binits{J.}},
\bauthor{\bsnm{{Wright}}, \binits{A.N.}},
\bauthor{\bsnm{{Hood}}, \binits{A.W.}}:
\byear{2020},
\batitle{{How Is Helicity (and Twist) Partitioned in Magnetohydrodynamic
  Simulations of Reconnecting Magnetic Flux Tubes?}}
\bjtitle{\apj}
\bvolume{898},
\bfpage{1}.
\doiurl{https://doi.org/10.3847/1538-4357/ab9c2a}.
\adsurl{2020ApJ...898....1T}.
\end{barticle}
\endbibitem

\bibitem[\protect\citeauthoryear{{Titov} and
  {D{\'e}moulin}}{1999}]{paper:TitovDemoulin1999}
\begin{barticle}
\bauthor{\bsnm{{Titov}}, \binits{V.S.}},
\bauthor{\bsnm{{D{\'e}moulin}}, \binits{P.}}:
\byear{1999},
\batitle{{Basic topology of twisted magnetic configurations in solar flares}}.
\bjtitle{\aap}
\bvolume{351},
\bfpage{707}.
\adsurl{1999A&A...351..707T}.
\end{barticle}
\endbibitem

\bibitem[\protect\citeauthoryear{{van~Rossum} and
  {de~Boer}}{1991}]{vanRossum1991}
\begin{barticle}
\bauthor{\bsnm{{van~Rossum}}, \binits{G.}},
\bauthor{\bsnm{{de~Boer}}, \binits{J.}}:
\byear{1991},
\batitle{{Interactively Testing Remote Servers Using the Python Programming
  Language}}.
\bjtitle{CWI Quarterly}
\bvolume{4},
\bfpage{283}.
\end{barticle}
\endbibitem

\bibitem[\protect\citeauthoryear{{Zuccarello}, {Aulanier}, and
  {Gilchrist}}{2015}]{paper:Zuccarelloetal2015}
\begin{barticle}
\bauthor{\bsnm{{Zuccarello}}, \binits{F.P.}},
\bauthor{\bsnm{{Aulanier}}, \binits{G.}},
\bauthor{\bsnm{{Gilchrist}}, \binits{S.A.}}:
\byear{2015},
\batitle{{Critical Decay Index at the Onset of Solar Eruptions}}.
\bjtitle{\apj}
\bvolume{814}.
\doiurl{https://doi.org/10.1088/0004-637X/814/2/126}.
\adsurl{2015ApJ...814..126Z}.
\end{barticle}
\endbibitem

\end{thebibliography}

\end{article}
\end{document}